\documentclass[11pt]{article}

\usepackage{amssymb}
\usepackage{epsfig}
\usepackage[latin1]{inputenc}
\usepackage{color}
\usepackage{graphics}
\usepackage{graphicx}
\usepackage{amsthm}
\usepackage{amsmath}
\usepackage{amsfonts}

\numberwithin{equation}{section}

\newcommand{\beq}{\begin{equation}}
\newcommand{\eeq}{\end{equation}}
\newcommand{\ber}{\begin{eqnarray}}
\newcommand{\eer}{\end{eqnarray}}
\newcommand{\bea}{\begin{align}}
\newcommand{\ea}{\end{align}}
\newcommand{\bes}{\begin{split}}
\newcommand{\es}{\end{split}}

\begin{document}

\title{ Coulomb integrals and conformal blocks \\ in the $AdS_3$-WZNW model}

\author{Sergio~M.~Iguri\footnote{e-mail: siguri@iafe.uba.ar} \, and 
Carmen~A.~N\'u\~nez\footnote{e-mail: carmen@iafe.uba.ar}}

\date{\small Instituto de Astronom\'{\i}a y F\'{\i}sica del Espacio
  (CONICET-UBA)
\\ C.~C.~67, Suc.~28, 1428 Buenos Aires, Argentina \\ and \\
  Departamento de 
F\'\i sica, F.C.E.~y~N., Universidad de Buenos Aires \\ Ciudad
  Universitaria, 
Pabell\'on 1, 1428 Buenos Aires, Argentina}

\maketitle

\begin{abstract}
We study spectral flow preserving four-point correlation functions in
the 
$AdS_3$-WZNW model using the Coulomb gas method on the sphere. 
We present a multiple integral realization of the conformal
blocks and explicitly compute 
amplitudes involving operators with  quantized values of
the sum of 
their spins, {\em i.e.}, requiring an integer number of screening
charges 
of the first kind.
The result is given 
as a sum over the
independent configurations of screening contours yielding a monodromy
invariant expansion in powers of the worldsheet moduli. We then examine
the factorization limit and show that the leading
terms in the sum can be identified, in the semiclassical limit, 
 with products of spectral flow conserving
three-point functions.
These terms can be
rewritten as 
the $m$-basis version of 
the integral expression obtained 
by J.~Teschner from a
postulate for the 
operator product expansion of normalizable states in the $H_3^+$-WZNW
model. 
Finally, we determine the equivalence between the factorizations of
a particular set of four-point functions
into products of two three-point
functions either preserving
or violating spectral flow number conservation.
 Based on this analysis 
we argue that the expression for the amplitude as an
 integral over the spin of the intermediate operators
holds 
beyond the semiclassical regime, thus
corroborating  that spectral flow conserving correlators
in the
$AdS_3$-WZNW
model  are related by analytic continuation to correlation functions in 
 the $H_3^+$-WZNW model.
\end{abstract}

\newpage

\tableofcontents

\bigskip

\section{Introduction}

In this article we continue the task, started in \cite{in}, of
computing correlation 
functions in the $AdS_3$-WZNW model within the Coulomb gas approach. 
In our first paper we used the Wakimoto representation to evaluate
both the spectral 
flow conserving and  violating three-point functions
of this theory on the sphere
and we showed that a proper analytic continuation to non-integer
numbers of screening 
operators gives amplitudes in full agreement with the exact results
previously obtained 
in \cite{tesch1, tesch3, mo3}. Here we  focus on the spectral flow
conserving  
four-point functions.

Among the many motivations for considering these correlators we can
mention their 
applications to string theory on $AdS_3$ and the $AdS$/CFT
correspondence as well 
as further examining the $AdS_3$-WZNW  model as a
prototype of non-rational 
conformal field theory (CFT) with affine Lie algebra symmetry, having
close connections 
with  Liouville theory  as well as with two and three dimensional gravity. 
Renewed interest in the
study of the conformal blocks  originates in
the recent developments presented in \cite{alday} (see also
\cite{morozov}) where it is conjectured that the conformal blocks
of  Liouville theory 
are related to
 the Nekrasov partition function of a certain class of
${\cal N}=2$ superconformal field theories \cite{nekrasov, nekoko}.

Most of what is known about the $AdS_3$-WZNW model is based on the
analytic continuation 
from its better understood Euclidean counterpart. The solution of the
$H_3^+$-WZNW model on 
the sphere was achieved in \cite{tesch1, tesch3} through a
generalization of the chiral 
bootstrap program and more recently it was further explored by means
of its relation to 
Liouville theory \cite{rt}. Specifically, it was proved that arbitrary
correlation 
functions  on the sphere can be expressed in
terms of 
correlators in Liouville theory which is so far the best understood
example of 
non-rational CFT \cite{zamo, teschl,naka}. 
However, there are many subtleties in the analytic continuation
relating the $H_3^+$ and the $AdS_3$ models. In particular,
the spectral flow automorphism of the latter is a highly
non-trivial feature determining a
fundamental problem in the application of the bootstrap program. Namely, while
 the contributions of primary states 
to the operator product expansion (OPE) 
in the
$H_3^+$ model are sufficient to complete this
program, the descendants not being
strictly necessary,  the spectral flow operation generates new representations
in which the conformal weights are not bounded below,
and thus 
the implementation of the bootstrap approach in the $AdS_3$ model 
requires a better understanding of the interplay between different 
spectral flow  sectors and possibly the explicit computation of 
 correlation functions involving affine descentant fields.

Observables in the $SL(2, \mathbb C)/SU(2)\equiv H_3^+$-WZNW model
are given by 
normalizable functions on the hyperbolic space, having the following
form in terms of 
Poincar\'e coordinates $(\phi, \gamma, \overline\gamma)$:
\begin{equation}
\Phi^j(x,\overline x| z, \overline z)=\frac{1+2j}{\pi}\left (
e^{-\sqrt{\frac 1{2(k-2)}}\phi}+
|\gamma - x|^2 e^{\sqrt{\frac 1{2(k-2)}}\phi}\right )^{2j},
\label{sf}
\end{equation}
where $k$ is the affine level of the algebra, $j$ labels the spin of
the state, 
$(z,\overline z)$ are the worldsheet coordinates and $(x,\overline x)$
keep track of the $SL(2, \mathbb C)$ quantum numbers. The Hilbert space of this
theory is 
the direct sum of its affine representations with primary states
having conformal weights
given by 
$\Delta_j=j(j+1)\rho$, where  $j \in {\mathcal P}^+ \equiv -1/2 +
i{\mathbb R}_{>0}$ 
and  $\rho=-(k-2)^{-1}$ \cite{gawe}.

Two- and three-point functions involving these operators were obtained
in 
\cite{tesch1, tesch3} solving the differential equations satisfied by
degenerate fields of admissible representations. 
Based on a proposal for the OPE of
normalizable primary 
fields and on a rigorous treatment in the mini-superspace
approximation \cite{tesch2}, 
an expression for the four-point function  was also presented in \cite{tesch1, 
tesch3}.
It involves an integral over a continuous family of solutions of the 
Knizhnik-Zamolodchikov (KZ) equation, namely,
\ber
\langle\Phi^{j_1}(0)\Phi^{j_2}(x, \overline x|z, \overline
z)\Phi^{j_3}(1) 
\Phi^{j_4}(\infty)\rangle = \int_{{\mathcal P}} dj \, D(j_1,j_2,j)
D(j_4,j_3,j)B(j)^{-1}|{\cal F}_j(x|z)|^2,
\label{4ptt}
\eer
where ${\cal P}\equiv -1/2 + i{\mathbb R}$, $D(j_1,j_2,j)$ and
$D(j_4,j_3,j)$ 
are the structure constants,
$B(j)$ is the propagator of the intermediate state\footnote{These
  functions are
 explicitly given below in (\ref{citprev1}) and (\ref{97}).} and 
the conformal blocks have the following expansion:
\begin{equation}
{\cal F}_j(x|z)=z^{\Delta_j-\Delta_{j_1}-\Delta_{j_2}}x^{j-j_1-j_2}
\sum_{n=0}^\infty f_n(x)z^n.
\label{exp}
\end{equation}
Substituting this expression into the KZ equation it is found that
$f_0(x)$ 
obeys the hypergeometric equation  so that, after imposing
monodromy invariance, 
it is univocally determined\footnote{Actually, there are two
  well-defined independent contributions 
to the conformal blocks related by the  reflection 
symmetry and monodromy invariance requires including both of them.
The extension of the
domain of 
integration from ${\cal P}^+$ to ${\cal P}$
allows to keep only one of these contributions.}. 
All other $f_n(x)$ can
be iteratively 
computed as stated in \cite{tesch3}.

Equation (\ref{4ptt}) holds for operators with spins in the following domain:
\begin{eqnarray}
\left\{
\begin{array}{ll}
|{\mbox{Re}}(j_1+j_2+1)|<\frac{1}{2}, \qquad & |{\mbox{Re}}(j_3+j_4+1)|
<\frac{1}{2}, \\
|{\mbox{Re}}(j_1-j_2)|<\frac{1}{2}, \qquad & |{\mbox{Re}}(j_3-j_4)|
<\frac{1}{2}.
\end{array}
\right.
\nonumber
\end{eqnarray}
For other values of the spins there are poles in the integrand that
hit the contour of 
integration and the four-point function must be defined by analytic
continuation. 
Crossing symmetry was shown to follow from similar properties of a
related five-point 
function in Liouville theory \cite{tesch4}. The amplitude (\ref{4ptt})
was further 
studied  in the context of string theory on $AdS_3$ in \cite{mo3}
where, after 
integrating over the moduli space of the worldsheet, it was written as
a sum of 
products of three-point functions summed over intermediate states
lying in the 
 physical spectrum.

The Hilbert space of the $AdS_3$-WZNW model \cite{mo1} is very
different from that of 
the Euclidean model. It decomposes into  direct products of the
normalizable 
continuous and highest-weight discrete  representations of the
universal 
cover of the affine $SL(2,{\mathbb R})$ algebra and their spectral flow images,
 namely, $\hat {\cal
  C}_j^{\alpha,w} 
\otimes \hat {\cal C}_j^{\alpha,w}$ with $j \in {\cal P}^+$ and
$\alpha\in [0,1)$, and 
$\hat {\cal D}_j^{-,w} \otimes \hat {\cal D}_j^{-,w}$ with $-(k-1)/2 <
  j < -1/2$. 
The spectral flow parameter $w$ is an integer number. All the states
in the spectrum, 
except those lying in the unflowed continuous representations,
correspond to 
non-normalizable operators in
the $H_3^+$-WZNW model. In order to deal with highest-weight as well as
spectral flowed 
representations it is convenient to work in a basis where the
generators 
$J_0^3,\overline J_0^3$ are diagonalized. This is the so-called $m$-basis.
 Unflowed
operators in the $m$-basis are related to (\ref{sf}) through the
following integral 
transform:
\ber
\Phi^{j}_{m,\overline m}(z,\overline z)=\int d^2x \, x^{j-m}\overline
x^{j-\overline m}\Phi^{-1-j}(x, \overline x|z,\overline z),
\label{it}
\eer
where $m$, $\overline m$ represent the eigenvalues of $J_0^3$,
$\overline J_0^3$, 
respectively, and $m-\overline m\in \mathbb Z$. States 
 in $\hat {\cal
  D}_j^{-,w}$ ($\hat {\cal C}_j^{\alpha, w}$)
have $m,\overline{m}=j-{\mathbb N}_0$
($m,\overline{m}=\alpha +{\mathbb Z}$). The spectral flow images of the
primary states are obtained from
(\ref{it}) acting with the spectral flow operators $\Phi^{-k/2}_{\pm
  k/2, \pm k/2}$ \cite{mo3, mo1} and they have conformal weight
$\hat\Delta_{j,m,w}=
\Delta_j-mw-kw^2/4$. 

Definite expressions for two- and three-point correlation functions of
unflowed 
operators were given in \cite{satoh} performing the integral transform
from the 
$x$-basis results of the $H_3^+$-WZNW model and analytically continuing the
kinematical parameters. The
accuracy of the 
analytic continuation is supported by the fact that it leads to the
well-known fusion 
rules for admissible representations \cite{tesch1} and to the
classical tensor products 
of representations
of $SL(2,{\mathbb R})$ \cite{satoh}.

Concerning the applications to string theory, this analytic
continuation was the 
starting point for a physical interpretation of the worldsheet
correlation functions in 
terms of correlators in the boundary CFT and for the analysis of the
factorization of 
four-point functions involving unflowed short string states in
\cite{mo3}. Amplitudes 
involving spectral flowed operators were evaluated
transforming to the $m$-basis the two- and three-point functions of
the $H_3^+$-WZNW 
model and acting with the  spectral flow operator. This
process was applied, 
in particular, to obtain the $w=1$ three-point function from a special
four-point 
function containing one spectral flow operator\footnote{A related
  computation was 
performed within the Coulomb gas formalism in \cite{in}.}. The problem
with applying 
this procedure to (\ref{4ptt}) is that the KZ equation implies the
existence of 
singularities at $z=0,\,1,\,x$ and $\infty$, which together with
monodromy invariance 
require that the amplitude behaves as
\ber
\langle\Phi^{j_1}(0)\Phi^{j_2}(x, \overline x|z, \overline z)\Phi^{j_3}(1)
\Phi^{j_4}(\infty)\rangle \sim |z-x|^{2(k+j_1+j_2+j_3+j_4)},\nonumber
\eer
and, thus, the expansion in (\ref{exp}) converges, in principle, only
for $|z|<|x|$. 
Integrals (\ref{it}) transforming to the $m$-picture can be done
either term by term 
in the expansion in powers of $z$ or with the full correlator obtained
by summing all 
the descendant contributions, but it is not clear that
summation and 
integration will commute or that the sum of integrals over $x$ will converge.

In the sequel we  present an independent derivation of the four-point
function in 
the $AdS_3$-WZNW model 
directly in the $m$-basis. This basis has the advantage that
correlators of fields with 
different amount of spectral flow can be treated 
simultaneously, {\em i.e.},
all $w$-conserving 
amplitudes are the same except for a known factor depending on the
insertion points of 
the vertex operators. We use the Coulomb gas method, which
provides a well defined 
framework within which it should be possible to address this question.

Unlike the successful applications to the minimal models \cite{dotfat}
and the 
$SU(2)$-WZNW model for operators with half-integer spins \cite{dot2},
the scope of the 
background charge method in theories with continuous sets of primary
fields appears to 
be limited because they necessarily require non-integer numbers of
screening operators. 
The basic difficulty in going away from half integer $SU(2)$-like
spins is that one no 
longer deals with degenerate fields satisfying null vector equations. 
A related problem arose in the evaluation of amplitudes involving
operators with 
rational spins in admissible representations of the $SU(2)$-WZNW model
\cite{dot2}, 
which could not be accomplished due to the necessity of considering
screening currents 
with rational powers of the ghost fields and the related ambiguity arising in
the analytic 
continuation to non-integer numbers of screening charges\footnote{See
  also 
\cite{petkova,
rasmu, andreev} for alternative approaches using free
field 
representations.}. Nevertheless, the formalism has played an important
role in 
the resolution of Liouville theory where an analytic continuation for
the three-point 
function was originally defined in \cite{goulian} (see also
\cite{zamo, dot3, schn}). Similarly, 
it was shown that the multiple Coulomb integrals define the residues
of the  
on-mass-shell three-point functions not only in Liouville theory but also in 
Toda CFT
\cite{fatlit}. More recently, the suggestion in \cite{alday} relating
the conformal blocks in Liouville theory and Nekrasov´s partition 
functions revives the longstanding 
idea that all conformal field theories can be
effectively described in the free field formalism \cite{morozov2}
because both Dotsenko-Fateev integrals and Nekrasov's functions 
provide a basis for generalized hypergeometric integrals.

In the case of the $AdS_3$-WZNW model full agreement was
found in 
\cite{in}\footnote{See \cite{becker2, gn} for previous related work.}
among the exact 
three-point functions,
both preserving 
and violating $w$-number conservation, 
 and those computed via the free field approach
for generic values of $j$. The
analytic 
continuation to non-integer numbers of screening charges was performed
in \cite{in} by 
noticing that the coefficients in the discrete sums arising from the
contractions of the
 $\beta$-$\gamma$ ghost fields can be written in terms of hypergeometric functions
which have to be supplemented with an extra contribution determined by
monodromy 
invariance. Defining an unambiguous analytic continuation procedure
should open up the 
possibility of studying arbitrary correlation functions using the
Coulomb gas picture 
for general spins and for any real value of the algebra level. In this
paper we move one 
step forward and extend the techniques developed in our first work to
address the 
computation of $w$-conserving four-point functions. In particular, we
show that there 
is an alternative representation of the discrete sums arising from the
monodromy 
invariant combination of chiral and anti-chiral conformal blocks in terms of
an integral reproducing the $m$-basis expression which is obtained
after applying the 
transformation (\ref{it}) to (\ref{4ptt}).

As mentioned above, (\ref{4ptt}) was obtained from the OPE of
normalizable states in the $H_3^+$-WZNW model applying the
factorization ansatz. 
But
the OPE proposed in \cite{tesch1, tesch3} would yield an incorrect
zero answer if used 
to compute, for example, $w$-violating three-point functions 
in the $AdS_3$ model; in other words, relaxing the
semiclassical approximation in the Lorentzian
model
is more elusive
 than in the Euclidean one. Indeed, it was argued in \cite{wc} 
(see also \cite{ribault}) that a modified OPE should be considered in
 the former
 theory including both $w$-preserving and non-preserving
structure 
constants and it was pointed out that this prescription gives fusion
rules of physical 
states consistent with the spectral flow symmetry and determining the
closure of the 
operator algebra on the Hilbert space of the theory. Consequently, the
factorization 
ansatz would lead to a modified expression for the four-point
functions containing both 
sets of structure constants. However, relying on a plausible but
hypothetical identity 
between two sets of four-point functions, it was shown in \cite{wc}
that both channels 
give equivalent contributions for certain $w$-conserving amplitudes
and it was argued 
that this must also be the case for all $w$-conserving four-point
functions. Here we complete 
the proof of that identity using the Coulomb gas approach. Actually we
show that the Coulomb integral realizations of these 
two sets of amplitudes agree, thus providing new evidence to support
the claim in 
\cite{wc}. This  also allows us to conjecture that the results
obtained for the four-point functions in the semiclassical limit
hold for generic affine level.

The paper is organized  as follows. In section 2 we compute spectral
flow conserving 
four-point functions using the Coulomb gas method
for spin configurations requiring an integer
number of screening 
operators. In section 3 we examine the
factorization limit 
and    show that 
the leading terms in the
expansion of 
the amplitude in powers of the worldsheet moduli can be written,
in the semiclassical regime, as the
$m$-basis version of 
Teschner's integral expression for the $H_3^+$-WZNW model. We also prove
an identity 
between two sets of four-point functions
which allows to show that the factorization into
spectral flow conserving or violating three-point functions give
equivalent 
contributions to these amplitudes
and to suggest that the results obtained for the analytic continuation 
hold for arbitrary affine level. A summary and discussions are
included in section 4. 
Some technical details and other computations are contained in the appendices.

\section{Coulomb gas computation of four-point functions}

In this Section we  evaluate $w$-conserving four-point functions
involving spectral 
flow images of primary fields in the $AdS_3$-WZNW model
using the Coulomb 
gas method.

Within this formalism, the relevant expectation values are of the form
(see \cite{in}):
\begin{eqnarray}
\label{corr5}
A_4^{w=0}\left[
\begin{matrix}
j_1,j_2,j_3,j_4 \cr  m_1, m_2, m_3, m_4 \cr 
w_1, w_2, w_3, w_4\cr \end{matrix}\right] =  \left\langle
\prod_{i=1}^4{V}^{j_i,w_i}_{m_i,\overline{m}_i}(z_i,
\overline{z}_i) \prod_{a=1}^{N_+}\eta^+(\zeta^+_a, \overline \zeta^+_a)
\prod_{b=1}^{N_-}\eta^-(\zeta^-_b, \overline \zeta^-_b)
\mathcal{Q}_1^{s_1} \mathcal{Q}_2^{s_2}
\right\rangle,
\end{eqnarray}
where the vertices are given by
\begin{eqnarray}
V_{m,\overline m}^{j,w}(z, \overline z)= \left [e^{(j-m-w)u(z)}
e^{-i(j-m)v(z)} \times c.c.\right ]
e^{\sqrt{\frac 2{k-2}}\left(
  j + \frac{k-2}2 w  \right) \phi(z,\bar z)},
\label{vj}
\end{eqnarray}
the screening operators of the first and second kind are,       
respectively, 
\begin{eqnarray}
\label{screech}
\mathcal{Q}_1=\int  d^2y ~ \left [
\partial v(y)
e^{-u(y)+iv(y)}\times c.c.\right ]
e^{-\sqrt{\frac 2{k-2}}\phi(y,\overline y)}\, , \nonumber
\end{eqnarray}
and
\begin{eqnarray}
\mathcal{Q}_2=\int  d^2y ~
\left [\partial v(y)
e^{-u(y)+iv(y)}\times c.c.\right ]^{k-2}
e^{-\sqrt{2(k-2)}\phi(y, \overline y)}\, ,\nonumber
\end{eqnarray}
and the spectral flow vertices act as picture changing operators for
the spectral 
flow sectors and have the following form:
\begin{equation}
\eta^+(\zeta, \overline \zeta)= \frac{1}{\pi \Gamma(0)}
\left [e^{(k-2)u(\zeta)}
 e^{-i(k-1)v(\zeta)}\times c.c.\right ]
e^{\sqrt{2(k-2)}\phi(\zeta, \overline\zeta)},\nonumber
\end{equation}
\begin{equation}
 \eta^-(\zeta,
\overline\zeta)
=\frac{1}{\pi \Gamma(0)} \left[ e^{iv(\zeta)}\times c.c.\right].
\nonumber
\end{equation}
These spectral flow operators were introduced in \cite{in} where it
was proved that 
they reproduce, when inserted into multi-point 
amplitudes, the
prescription proposed
in \cite{zz} and applied in \cite{mo3}
to compute correlators involving spectral flowed states.

Here and thereafter ``$c.c.$'' indicates that all the variables have to be
replaced by the barred ones.

The vertex operators (\ref{vj}) are related, in the unflowed case and
in the 
large-$\phi$ limit, to those defined by (\ref{it}) through
 \begin{eqnarray}
\Phi^{j}_{m,\overline m}(z, \overline z)=V^{j}_{m,\overline m}(z, 
\overline z)
+
 B(-1-j)c^{-1-j}_{m,\overline m}
V^{-1-j}_{m,\overline m}(z, \overline z),
\label{vc}
\end{eqnarray}
where
\ber
B(j)=-\frac{1+2j}{\pi}\,\nu^{1+2j}\frac{\Gamma(1-\rho(1+2j))}
{\Gamma(1+\rho(1+2j))}, 
\qquad\qquad \nu=\pi\,\frac{\Gamma(1+\rho)}{\Gamma(1-\rho)},
\label{citprev1}
\eer
and
\begin{eqnarray}
c^{j}_{m,\overline m}=
\pi\gamma(1+2j)\,
\frac{\Gamma(-j-m)\Gamma(-j+\overline
  m)}{\Gamma(1+j-m)\Gamma(1+j+\overline m)}, 
\qquad\qquad \gamma(x)=\frac{\Gamma(x)}{\Gamma(1-x)}.\nonumber
\end{eqnarray}

The charge asymmetry conditions for the free field expectation values
in (\ref{corr5}) are given by
\begin{eqnarray}
\sum_{n=1}^4 j_n=s_1+(k-2)\left[s_2-\frac{N_++N_-}{2}\right]-1\, ,
\label{cj}
\end{eqnarray}
\begin{eqnarray}
\sum_{n=1}^4 m_n = \sum_{n=1}^4 \overline{m}_n = \frac k2\,(N_+-N_-),\nonumber
\end{eqnarray}
\begin{eqnarray}
\sum_{n=0}^4 w_n=N_--N_+.
\label{consl3}
\end{eqnarray}
As expected, for $w$-conserving amplitudes we must take $N_+=N_-$.

Notice from (\ref{cj}) that ${\cal Q}_2$, which only makes sense for
$k\in{\mathbb N}_{>2}$, screens exactly the charge carried by a couple
of spectral flow 
operators of each kind. On the other hand, since correlation functions
in the $m$-basis 
depend on the total $w$-number and the only change in $w$-conserving
correlators 
involving states in different spectral flow sectors is contained in
the powers of the 
worldsheet coordinates, assuming $w_i=0$ for $i=1,\cdots, 4$, does not
imply any loss 
of generality and we can further take $N_+=N_-=0$.

Only correlators with vertices requiring positive integer numbers of
 screenings, namely,
 $s_1,s_2\in {\mathbb N}_0$, can be directly computed in this
 formalism. Correlation 
functions involving operators in continuous representations or their
 spectral flow
images cannot be considered at once in this picture because one cannot
 choose the 
imaginary parts of the spins in both terms of (\ref{vc}) so that they
 add up to zero 
in all the terms of the full amplitude. Instead, the second term of
 the vertex operators 
creating states in discrete representations vanishes\footnote{See
 \cite{hs} for a 
discussion on the different asymptotic behaviours of operators in 
highest-weight or
continuous representations.} and therefore, the number of charges needed 
to screen
four operators in discrete series turns out to be negative, due to the
 values of the spins.
Negative powers of screenings have been considered in Liouville theory
\cite{dot3, schn} and it was shown that there exists a consistent
extension of the formalism to deal with this situation. Alternatively, one
can use the reflection symmetry 
in order to work with positive numbers of screenings. In
 the sequel we  
adopt the latter option.
In conclusion, only certain states with particular spin values in
 discrete 
representations can satisfy  equation (\ref{cj}) and results for generic
configurations require analytic continuation.
Therefore, we can take $s_2=0$ without loosing generality, and thus 
$k\in {\mathbb R}_{>2}$.

Summarizing, in this section  we evaluate 
four-point correlation functions involving operators in the 
unflowed principal discrete
representations with $s_2=0$, $N_+=N_-=0$,
namely,
\ber
A_4^{w=0}\left[
\begin{matrix}
j_1,j_2,j_3,j_4 \cr  m_1, m_2, m_3, m_4 \cr \end{matrix}\right] 
\equiv \Gamma(-s)
 \left\langle
{V}^{j_1,w_1=0}_{m_1,\overline{m}_1}(0){V}^{j_2,w_2=0}_{m_2,\overline{m}_2}(z,
\overline{z})
 {V}^{j_3,w_3=0}_{m_3,\overline{m}_3}(1){V}^{j_4,w_4=0}_{m_4,\overline{m}_4}
(+\infty)
\mathcal{Q}_1^{s}
\right\rangle,
\label{very}
\eer
where  $s\equiv s_1=j_1+
\cdots+j_4+1 \in \mathbb{N}_0$.
The spectral flow labels in the arguments on the l.h.s. are omitted for
short\footnote{We shall explicitly write these labels in section 3,
  when the spectral flow numbers of the operators become relevant.},  
$w$  refers to the total spectral flow number of the
amplitude and global conformal
invariance 
was used in order to set $z_1,\overline{z}_1=0$, $z_2=z$,
$\overline{z}_2=\overline{z}$, 
$z_3,\overline{z}_3=1$ and $z_4,\overline{z}_4=\infty$.
The factor $\Gamma(-s)$ arises from the integration of the zero modes of 
$\phi(z,\overline z)$; by abuse of notation, we also denote the vertex 
operators
  as $V_{m,\overline m}^{j,w}$ 
after performing this integration.

The corresponding field contractions give:
\begin{eqnarray}
\label{acasi}
A_4^{w=0}\left[
\begin{matrix}
j_1,j_2,j_3,j_4 \cr  m_1, m_2, m_3,m_4 \cr \end{matrix}\right] =
\Gamma(-s) 
|z|^{4 j_1 j_2 \rho} |1-z|^{4 j_2 j_3 \rho}
 \int [dy] \prod_{i=1}^s
|y_i|^{-4j_1\rho} |z-y_i|^{-4j_2\rho} \nonumber \\
\times |1-y_i|^{-4j_3\rho}
\prod_{j>i}^s |y_i-y_j|^{4\rho} \left[
\frac{1}{P} \,
\partial_1
\cdots \partial_s P \times c.c. \right],
\end{eqnarray}
where $[dy]$  is a shorthand  for $\prod_{i=1}^s d^2 y_i$, 
$\partial_i\equiv \partial/\partial y_i$, and
\begin{eqnarray}
\label{ppp}
P=\prod_{i=1}^sy_i^{-j_1+m_1}(z-y_i)^{-j_2+m_2}(1-y_i)^{-j_3+m_3}
\prod_{j>i}^s(y_i-y_j)
\end{eqnarray}
is the contribution  from the (holomorphic) ghost system. 
Recall that the field $\phi$ and the  free bosons $u$ and
$v$, which 
bosonize the usual $\beta$-$\gamma$ ghost system, have propagators:
\begin{equation}
\langle u(z)u(w)\rangle = \langle v(z)v(w)\rangle =\langle
\phi(z)\phi(w)\rangle =
-{\rm log}(z-w),\nonumber
\end{equation}
and similar expressions hold for the anti-holomorphic components.

In the next subsection we compute the ghost contributions and
then we proceed to the evaluation of the Coulomb integrals.

\subsection{Contributions from the ghost system}

It is convenient to recall the definition of the Vandermonde
determinant:
\begin{equation}
\prod_{i=1}^s\prod_{j>i}^s(y_i-y_j)=\det\left(y_i^{j-1}\right),
\nonumber
\end{equation}
and use it to rewrite  (\ref{ppp}) as follows:
\begin{eqnarray}
P = \det \left[(z-y_i)^{-j_2+m_2}(1-y_i)^{-j_3+m_3}
y_i^{j-1-j_1+m_1}\right].\nonumber
\end{eqnarray}
Taking derivatives as
\begin{eqnarray}
\partial_1 \cdots \partial_s P & = & \det\left\{
  \partial_i\left [(z-y_i)^{-j_2+m_2}
(1-y_i)^{-j_3+m_3} y_i^{j-1-j_1+m_1} \right ]\right \} \nonumber\\
& = &
 \left[\prod_{i=1}^s y_i^{-j_1+m_1-1}
(z-y_i)^{-j_2+m_2-1} (1-y_i)^{-j_3+m_3-1}\right] \det\left[
  \sum_{l=0}^2 \ell_l^j(z)
y_i^{j-1+l} \right],
\nonumber
\end{eqnarray}
we can write
\begin{eqnarray}
\label{refere3}
\frac{1}{P} \, \partial_1 \cdots \partial_s P = \left[\prod_{i=1}^s
y_i^{-1}(z-y_i)^{-1}
(1-y_i)^{-1}\right] \frac{\det\left[ \sum_{l=0}^2 \ell_l^j(z) y_i^{j-1+l}
    \right]}
{\det\left(y_i^{j-1} \right)}\, ,
\end{eqnarray}
where we have introduced:
\begin{eqnarray}
\left\{
\begin{array}{l}
\ell_0^j(z)= (j-1-j_1+m_1)z, \\
 \ell_1^j(z)= 1-j+j_1+j_2-m_1-m_2 + (1-j+j_1+j_3-m_1-m_3)z,  \\
\ell_2^j(z)= j-1-j_1-j_2-j_3+m_1+m_2+m_3 . \end{array} \right. 
\nonumber
\end{eqnarray}

 Notice that the entries of the matrix in the numerator
are three term polynomials in the screening variables with powers exceeding in
$2$, $1$ and $0$ units the corresponding powers of the entries of the matrix
in the denominator. Using the multilinearity of the determinants and
performing all the 
 distributions we get
\begin{eqnarray}
\det\left[ \sum_{l=0}^2 \ell_l^j(z) y_i^{j-1+l} \right] &=&
\sum_{\lambda \in \left[0,2
\right]^s} \det\left[\ell_{\lambda_{s+1-j}}^j(z)
  y_i^{j-1+\lambda_{s+1-j}}\right] 
\nonumber
\\
&=& \sum_{\lambda \in \left[0,2\right]^s} \left[ \prod_{j=1}^s
\ell_{\lambda_{s+1-j}}^j(z)
\right] \det\left(y_i^{j-1+\lambda_{s+1-j}}\right), \nonumber
\end{eqnarray}
 where the sum index
$\lambda$ is the $s$-tuple whose components give the excedent in power of the
matrix elements
with respect to those of the Vandermonde determinant, in the inverse order.
Thus the quotient of determinants in (\ref{refere3}) looks like the one 
defining 
 Schur polynomials,
except for the fact that $\lambda$ is not a partition but an
$s$-tuple with
entries taking the values
$0$, $1$ and $2$. Nevertheless, we shall show that
it is possible to  rewrite (\ref{refere3}) so as to sum only over
partitions.

The $s$-tuples of the form
$(\dots,0,1,\dots)$ or $(\dots,1,2,\dots)$  give no contribution to
the quotient in
 (\ref{refere3}) since the determinant in the numerator vanishes, and thus the
sum is over partitions except for the
case of  $s$-tuples of the form
$(\dots,0,2,\dots)$. But in this case neither  the $s$-tuples of
the form
$(\dots,0,2,2,\dots)$
nor those of the form $(\dots,0,0,2,\dots)$ contribute because, again, the
determinant in the numerator vanishes.
Consequently, only the following $s$-tuples
are relevant:
\begin{eqnarray}
\label{efec}
\lambda=(2,\dots,2,1,\dots,1,0,2,1,\dots,1,0,2,1,\dots,1,0,\dots,0)\, .
\end{eqnarray}

Since the interchange of two columns in a determinant only  changes
its overall sign, the quotient of determinants in (\ref{refere3}) associated
with the $s$-tuple (\ref{efec}) is equal to the Schur polynomial associated to
the partition $(2,\dots,2,1,\dots,1,0,\dots,0)$ up to a factor $(\pm 1)$ depending on the
number of times the subsequence ``$\dots,0,2,\dots$'' is replaced by ``$\dots,1,1,\dots$''
in (\ref{efec}) in order to obtain a partition.
This implies that we can actually write
\begin{eqnarray}
\frac{\det\left[ \sum_{l=0}^2 \ell_l^j(z) y_i^{j-1+l}
    \right]}{\det\left(y_i^{j-1} \right)}
= \sum_{\lambda} C_{\lambda}(z) s_{\lambda}(y_1,\dots,y_s),
\nonumber
\end{eqnarray}
where now the sum is over partitions of length $s$ and
entries $0, 1$ or 2.
These partitions are characterized by two integer numbers, say $n$ and $r$,
denoting the number of times the entries $2$ and $1$ appear, respectively.
 Let us write $C_{nr}(z)$ instead of $C_{\lambda}(z)$ and $s_{nr}
(y_1,\dots,y_s)$ instead of $s_{\lambda}(y_1,\dots,y_s)$. We then have:
\begin{eqnarray}
\frac{1}{P}\,\partial_1 \cdots \partial_s P = \left[\prod_{i=1}^s
  y_i^{-1}(z-y_i)^{-1}
(1-y_i)^{-1}\right]\sum_{n=0}^s \sum_{r=0}^{s-n} C_{nr}(z)
  s_{nr}(y_1,\dots,y_s), \nonumber
\end{eqnarray}
so that the four-point function (\ref{acasi}) may be rewritten as
\begin{eqnarray}
 A_4^{w=0}\left[
\begin{matrix}
j_1,j_2,j_3,j_4 \cr  m_1, m_2, m_3,m_4 \cr \end{matrix}\right] = 
\Gamma(-s) |z|^{4 j_1 j_2 \rho} |1-z|^{4 j_2 j_3 \rho}
\sum_{n,\overline{n}=0}^s\sum_{r,\overline{r}=0}^{s-n}
\left[ C_{nr}(z) \times c.c. \right]
{\cal J}_{nr,\overline
n\overline r}(z,\overline z), \nonumber
\end{eqnarray}
where ${\cal J}_{nr,\overline n\overline r}(z,\overline z)$ are
the following generalized Selberg complex integrals:
\begin{eqnarray}
{\cal J}_{nr,\overline n\overline r}(z,\overline z)=\int [dy]
\prod_{i=1}^s |y_i|^{-4j_1
\rho-2} |z-y_i|^{-4j_2\rho-2}
\times |1-y_i|^{-4j_3\rho-2} \prod_{i<j}^s |y_i-y_j|^{4\rho}\nonumber\\
\times
s_{nr}(y_1,\dots,y_s)
s_{\overline{n}\overline{r}}(\overline{y}_1,\dots,\overline{y}_s).\label{212b}
\end{eqnarray}

Therefore, the problem has been reduced to two independent calculations, 
namely, 
obtaining the coefficients $C_{nr}(z)$ and performing the computation
of the Coulomb 
integrals ${\cal J}_{nr,\overline n\overline r}(z,\overline{z})$. The
coefficients 
$C_{nr}(z)$ are computed in the Appendix A.1. They involve complicated 
hypergeometric-like expansions with polynomials as arguments (see
(\ref{pq})). 
In the following subsection we compute them
assuming that a 
highest-weight state is inserted at $z_1,\overline{z}_1=0$; many simplifications occur in
this case and this allows us to deal with the Coulomb integrals in subsection 2.3.

\subsection{Evaluation of $m$-dependent coefficients: one
highest-weight state}

Explicitly evaluating  the terms
contributing to  $C_{nr}(z)$ in (\ref{refere3}) for
different values of $n$ and $r$, it can be shown that
\begin{eqnarray}
C_{nr}(z) = (-1)^{s-n-r} z^{s-n-r} \, \frac{\Gamma(\alpha+1)}{\Gamma(\alpha
  -s+n+r+1)}
\frac{\Gamma(s-\alpha-\beta-\gamma)}{\Gamma(s-n-\alpha-\beta-\gamma)}\,d_{s-n}(z),
\label{cnp}
\end{eqnarray}
where we have defined
\begin{equation}
\left\{
\begin{array}{l}
\alpha=j_1-m_1, \\
\beta =j_2-m_2, \\
\gamma = j_3-m_3,
\end{array}
\right.
\nonumber
\end{equation}
 $d_{s-n}(z)$ is the determinant of the matrix
$(a_{ij}(z))_{i,j=1}^{s-n}$
with entries given by
\begin{equation}
\label{matrix}
a_{ij}(z)=\ell^{s-n-r+j}_{i-j+1}(z),
\end{equation}
and we are setting $a_{ij}(z)=0$ if $|i-j|>1$.

As we have mentioned, many simplifications occur if the operator inserted at
$z_1,\overline{z}_1=0$ creates a highest-weight state, so from
here on we assume  $j_1=m_1$.
In Appendix A.1 we  present the computation in full generality.

The coefficients $C_{nr}(z)$ vanish when $\alpha=0$
unless $r=s-n$ and in this case we have
\begin{eqnarray}
C_{n}(z) \equiv C_{nr=s-n}(z) = \frac{\Gamma(s-\beta-\gamma)}
{\Gamma(s-n-\beta-\gamma)}\,d^0_{s-n}(z),
\nonumber
\end{eqnarray}
where $d^0_{s-n}(z)$ is the determinant of the matrix
$(a_{ij}^0(z))^{s-n}_{i,j=1}$ with elements
$a^0_{ij}(z)=\ell^{j}_{i-j+1}(z)$.

Let us denote by $d^0_p(z)$, $p=1,2,\dots,s-n$,
the determinant of the matrix $(a^0_{ij}(z))_{i,j=1}^{p}$ formed by
the first $p$ rows and $p$ columns of
  $(a^0_{ij}(z))_{i,j=1}^{s-n}$. Notice that
$d^0_p(z)$ is a polynomial in $z$ of degree $p$ satisfying the
  following
recurrence formula:
\begin{equation}
d^0_p(z)=\ell_1^p(z) d^0_{p-1}(z) - \ell_2^{p-1}(z) \ell_0^p(z) d^0_{p-2}(z),
\nonumber
\end{equation}
or, more explicitly,
\begin{equation}
\label{recurr}
d^0_p(z)=[(1-p+\beta) + (1-j+\gamma)z] d^0_{p-1}(z) - (p-2-\beta-\gamma) (p-1) 
z
d^0_{p-2}(z),
\end{equation}
which follows from the fact that $(a^0_{ij}(z))_{i,j=1}^{p}$ is a tridiagonal
matrix.

The boundary conditions for this recurrence are: $d^0_1(z)=\ell_1^1(z)$ and
$d^0_2(z) = \ell_1^1(z) \ell_1^2(z) - \ell_2^1(z) \ell_0^2(z)$.

It can be inductively
proved that the solution of (\ref{recurr}) is given by
\begin{equation}
d^0_p(z)= \frac{\Gamma(\beta+1)}{\Gamma(-p)\Gamma(-\gamma)}
\sum_{t=0}^p  \frac{\Gamma(-p+t)
  \Gamma(-\gamma+t)}
{\Gamma(\beta - p + 1 + t)} \frac{z^t}{t!}.\label{sum}
\end{equation}

Finally, noticing that the sum over $t$ can be  freely
 taken to $\infty$,
we can write
\begin{equation}
d^0_{s-n}(z)= \frac{\Gamma(\beta+1)}{\Gamma(\beta - s + n + 1 )} \,
{}_2F_1\left [ \left .\begin{array}{c}
-s+n,-\gamma \\ \beta-s+n+1 \end{array} \right | z \right ],\nonumber
\end{equation}
and then,
\begin{eqnarray}
C_n=B_n \, {}_2F_1\left [ \left .\begin{array}{c}
-s+n,-j_3+m_3 \\ j_2-m_2-s+n+1 \end{array} \right | z \right ],
\label{sc}
\end{eqnarray}
where
\begin{equation}
B_n\equiv\frac{\Gamma(s-j_2-j_3+m_2+m_3)\Gamma(j_2-m_2+1)}
{\Gamma(s-n-j_2-j_3+m_2+m_3)
\Gamma(j_2-m_2-s+n+1)}.
\label{bn}
\end{equation}
On the other hand, since only partitions of the form
$(2,\dots,2,1,\dots,1)$ appear we can use
\begin{equation}
s_{n,s-n}(y_1,\dots,y_s)
= \left[ \prod_{i=1}^s y_i \right] \times
\alpha^s_{n}(y_1,\dots,y_s)\, ,
\nonumber
\end{equation}
where
\begin{equation}
\alpha_n^s(y_1,\cdots ,y_s) = \frac
1{n!(s-n)!}\sum_{\sigma_n}\prod_{i=1}^n
y_{\sigma_{n(i)}}\nonumber
\end{equation}
is an elementary symmetric polynomial and $\alpha_0^s=1$,
 to finally conclude that the four-point function involving one
highest-weight state is given by
\begin{eqnarray}
 A_4^{w=0}\left[
\begin{matrix}
j_1,j_2,j_3,j_4 \cr  j_1, m_2, m_3,m_4 \cr \end{matrix}\right]
~ ~ ~ ~ ~ ~ ~ ~ ~ ~ ~ ~ ~ ~ ~ ~ ~ ~ ~ ~ ~ ~ ~ ~ ~ ~ ~ ~ ~ ~  
~ ~ ~ ~ ~ ~ ~ ~ ~ ~ ~ ~ ~ ~ ~ ~ ~ ~ ~ ~ ~ ~ ~ ~ ~ ~ ~ ~ ~ ~ 
~ ~ ~ ~ ~ ~ ~ ~ ~ ~ ~ ~ ~ ~ ~ ~ ~ ~ ~ ~ ~ ~ ~ ~ ~ ~ ~ ~ ~ ~ ~ ~ ~ ~ ~ ~ ~ ~ ~ 
\nonumber \\
~ ~ ~ ~ ~ ~ ~ ~  = \Gamma(-s) |z|^{4 j_1 j_2 \rho} |1-z|^{4 j_2 j_3 \rho} 
\sum_{n,\overline{n}=0}^s
\left\{ B_n \, {}_2F_1\left [ \left .\begin{array}{c}
-s+n,-j_3+m_3 \\ j_2-m_2-s+n+1 \end{array} \right | z \right ]
\times c.c. \right\} \times
\mathcal{J}_{n,\overline{n}}(z,\overline z),
\label{4p1pm}
\end{eqnarray}
where
we have defined
\begin{eqnarray}
\label{integ}
\mathcal{J}_{n,\overline{n}}(z,
\overline{z}) =
 \int [dy] \prod_{i=1}^s |y_i|^{-4j_1\rho} |z-y_i|^{-4j_2\rho-2}
      |1-y_i|^{-4j_3\rho-2}
\prod_{i<j}^s |y_i-y_j|^{4\rho} \nonumber \\
\times ~\alpha^s_{n}(y_1,\dots,y_s)\,\overline{\alpha}^s_{\overline{n}}
(\overline{y}_1,\dots,
\overline{y}_s).
\end{eqnarray}

Notice that, other than in the explicit overall factor, the
$z$-dependence of (\ref{4p1pm})
appears in the hypergeometric function and both in the integrand and in
the integration domain
of ${\mathcal J}_{n,\overline n}(z,
\overline{z})$.
In the next section we  analyze this dependence in detail.

\subsection{Monodromy invariance and normalization}

According to the analysis in \cite{dotfat},
the integral ${\cal J}_{n,\overline{n}}(z,\overline{z})$
is given by the monodromy invariant combination of
chiral and antichiral conformal blocks as
\begin{eqnarray}
\label{21}
{\cal J}_{n,\overline{n}}(z, \overline z) = \sum_{l=0}^s
X_{l}^{n\overline{n}} I_{n}^{l}(z)
I_{\overline{n}}^{{l}}(\overline z)\, ,
\end{eqnarray}
where
\begin{eqnarray}
 I_{n}^l(z) = \int_{\Delta^{(1,\infty)}_{s-l}} \prod_{i=1}^{s-l} dy_i \int_{\Delta^{(0,z)}_{l}} \prod_{i=1}^l dy_{s-l+i} \prod_{i=1}^s y_i^{-2j_1\rho}\prod_{i<j}^s (y_i-y_j)^{2\rho} \prod_{i=1}^{s-l}
(y_i-z)^{-2j_2\rho-1}(y_i-1)^{-2j_3\rho-1}
  \nonumber \\
\times
\prod_{i=1}^{l} (z-y_{s-l+i})^{-2j_2\rho-1}
(1-y_{s-l+i})^{-2j_3\rho-1}
\alpha^s_{n}(y_1,\dots,y_s),
\nonumber
\end{eqnarray}
the integration domains being the simplex $\Delta^{(1,\infty)}_{s-l}
\equiv\{(y_1,\dots,y_{s-l}):1<y_{s-l}<\cdots<y_1<+\infty\}$ and 
$\Delta^{(0,z)}_{l}\equiv\{(y_{s-l+1},\dots,y_s):0<y_{s}<\cdots<y_{s-l+1}<z\}$.

The form (\ref{21}) is diagonal in $I_n^l(z)$ since these functions have
diagonal $s$-channel
monodromy (as we show below).
The coefficients $X_{l}^{n\overline{n}}$ are
determined from the requirement that the physical four-point function must
be monodromy invariant with respect to the analytic continuation over
$z$ and $\overline z$
around $z,\overline z=0$ and around $z,\overline z=1$.

Alternatively to $I_n^l(z)$, one may consider the following 
``unordered'' integrals:
\begin{eqnarray}
\label{17}
 J_{n}^l(z) 
&=& \int_{(1,\infty)^{s-l}} \prod_{i=1}^{s-l} dy_i \int_{(0,z)^{l}} 
\prod_{i=1}^l dy_{s-l+i} \prod_{i=1}^s y_i^{-2j_1\rho}\prod_{i=1}^{s-l}~
(y_i-z)^{-2j_2\rho-1}(y_i-1)^{-2j_3\rho-1} \nonumber \\
&& \times ~ \prod_{i=1}^{l} (z-y_{s-l+i})^{-2j_2\rho-1}
(1-y_{s-l+i})^{-2j_3\rho-1}~ \prod_{i<j}^s (y_i-y_j)^{2\rho}
 \alpha^s_{n}(y_1,\dots,y_s),
\end{eqnarray}
which are related to $I_n^l(z)$ as
\begin{eqnarray}
J_{n}^l(z) = \lambda_l(\rho) I_{n}^l(z), \nonumber
\end{eqnarray}
with
\begin{eqnarray}
&& \lambda_l(\rho) = \prod_{i=1}^{s-l} \frac{\sin(i \pi
  \rho)}{\sin(\pi \rho)}
\prod_{i=1}^{l} \frac{\sin(i \pi \rho)}{\sin(\pi \rho)}  \, . \nonumber
\end{eqnarray}
The symmetry of
$\alpha^s_{n}(y_1,\dots,y_s)$  under any permutation of its arguments renders
the proof of this statement  exactly as in \cite{dotfat}.
For given values of $n$, the functions (\ref{17}) with different
values of $l$ are mutually independent. They provide the integral
representation for the system of diagonal conformal blocks with
respect to the $s$-channel.

In order to prove that
the monodromy group around $z = 0$ acts diagonally on
the basis $\{J_n^l(z)\}$,
we perform the change of variables
$y_{s-l+q}=zu_q$
for $q=1,2,\dots,l$ and  get
\begin{eqnarray}
\label{27}
 J_{n}^l(z) = z^{-2lj_1\rho-2lj_2\rho+\rho l(l-1)}
 \int_{(1,\infty)^{s-l}} 
\prod_{i=1}^{s-l} dy_i \int_{(0,1)^{l}} \prod_{i=1}^l du_{i} 
\prod_{i=1}^{s-l} y_i^{-2j_1\rho} (y_i-z)^{-2j_2\rho-1} \nonumber \\
\times (y_i-1)^{-2j_3\rho-1} \prod_{i<j}^{s-l} (y_i-y_j)^{2\rho}
\prod_{q=1}^l u_q^{-2j_1\rho} (1-u_q)^{-2j_2\rho-1} (1-zu_q)^{-2j_3\rho-1}
\nonumber \\
\times  \prod_{q<p}^l (u_q-u_p)^{2\rho} \prod_{i=1}^{s-l} \prod_{q=1}^l
 (y_i-zu_q)^{2\rho} \alpha^s_{n}(y_1,\dots,y_{s-l},zu_1,\dots,zu_l)\, .
\end{eqnarray}
It is easy to see that $\alpha^s_{n}(y_1,\dots,y_{s-l},zu_1,\dots,zu_l)$ does
not change
the overall $z$-dependence of (\ref{27}) if $n=0,1,2,\dots,s-l$, but an
extra factor
$z^{l+n-s}$ appears if $n\ge s-l+1$.
However a rotation
around $z=0$ gives no additional phase factor since $l+n-s$ is an integer
number.
After extracting the $z$-dependence in (\ref{27}), the
integral is an
analytic
function, regular at $z=0$.
Consequently, a monodromy transformation around $z=0$  gives
\begin{eqnarray}
J_{n}^l(z) \rightarrow \exp[-2 l \pi i (2j_1+2j_2-l+1)\rho] \times
J_{n}^l(z)\, . \nonumber
\end{eqnarray}

The following integrals provide
an alternative basis for (\ref{integ}) \cite{dotfat}:
\begin{eqnarray}
 \widetilde{J}_{n}^l(z) = \int_{(-\infty,0)^{s-l}} \prod_{i=1}^{s-l} dy_i \int_{(z,1)^{l}} \prod_{i=1}^l dy_{s-l+i} \prod_{i=1}^{s-l} (-y_i)^{-2j_1\rho}
(z-y_i)^{-2j_2\rho-1}(1-y_i)^{-2j_3\rho-1} \nonumber \\
\times \prod_{i=1}^{l} y_{s-l+i}^{-2j_1\rho}(y_{s-l+i}-z)^{-2j_2\rho-1}
(1-y_{s-l+i})^{-2j_3\rho-1}\prod_{i>j}^s (y_i-y_j)^{2\rho}
\alpha^s_n(y_1,\dots,y_s). \nonumber
\end{eqnarray}
In this case we may 
 prove that this set  is a
canonical basis for the point $z=1$.
To this aim, it is convenient to
perform two changes of variables, first
$y_i \longrightarrow 1-y_{i}$ and then
$y_{s-l+q} \longrightarrow (1-z) u_q$,
for $q=1,2,\dots,l$. This gives:
\begin{eqnarray}
&&\widetilde{J}_{n}^l(z) = (1-z)^{-2lj_3\rho-2lj_2\rho-l+l(l-1)\rho}
\sum_{n'=0}^{n} (-1)^{n'} \left (\begin{array}{c} s-n'\\
  n-n'\end{array}\right )
 \int_{(1,\infty)^{s-l}} \prod_{i=1}^{s-l} dy_i \int_{(0,1)^{l}}
 \prod_{i=1}^l du_{i} 
\nonumber \\
&&~~~~\times \prod_{i=1}^{s-l} (y_i-1)^{-2j_1\rho} (y_i-(1-z))^{-2j_2\rho-1}
y_i^{-2j_3\rho-1} \prod_{i<j}^{s-l} (y_i-y_j)^{2\rho}  \prod_{q=1}^{l}
(1-(1-z)u_q)^{-2j_1\rho} u_q^{-2j_3\rho-1} \nonumber \\
&& ~~~~\times (1-u_q)^{-2j_2\rho-1} \prod_{q<p}^l
(u_q-u_p)^{2\rho}
\prod_{i=1}^{s-l} \prod_{q=1}^{l} (y_i-(1-z)u_q)^{2\rho} 
\alpha^s_{n'}(y_1,\dots,y_{s-l},(1-z)u_1,\dots,(1-z)u_l),\nonumber
\end{eqnarray}
where we have used the following identity
\begin{eqnarray}
\alpha^s_{n}(1-w_1,\dots,1-w_s) = \sum_{n'=0}^{n} (-1)^{n'}
\left (\begin{array}{c} s-n'\\
  n-n'\end{array}\right ) \alpha^s_{n'}(w_1,\dots,w_s), \label{223b}
\end{eqnarray}
which can be proved inductively.

In this case, if $n=0,1,\dots,s-l$, {\em i.e.}, $l+n\le s$, then $l+n' \le s$
and  there is no additional overall $(1-z)$-dependence coming from the
elementary 
symmetric polynomials, as we have seen.
If $l+n > s$, for  $l+n'>s$ an extra factor $(1-z)^{l+n'-s}$
should be considered. But, again, a loop around $z=1$ gives no
non-trivial phase
factor since $l+n'-s \in \mathbb{N}_0$. It follows that a monodromy
transformation around $z=1$ is given by
\begin{eqnarray}
\widetilde{J}_{n}^l(z) \rightarrow \exp[-2 l \pi i (2j_2+2j_3-l+1)\rho]
\times \widetilde{J}_{n}^l(z), \nonumber
\end{eqnarray}
and $\{\widetilde{J}_{n}^l(z)\}$ is, thus, a canonical basis for $z=1$.

Having checked that  $\{J_n^l(z)\}$ and $\{\tilde J_n^l(z)\}$
are canonical basis for the points $z=0$ and $z=1$,
respectively, the computation of the
factors $X_{l}^{n\overline{n}}$ follows as in \cite{dotfat}. Since they
do not depend on $n,\overline{n}$ we may write\footnote{Recall 
that these coefficients are defined up to an overall $l$-independent 
factor to be determined from the two-point function \cite{dotfat}. 
Here they are already normalized.}
\begin{eqnarray}
\label{59}
X_{l}\equiv X_l^{n\overline n}=\frac{1}{s!}\prod_{i=1}^{s-l}
\frac{\sin(i \pi \rho)}
{\sin(\pi \rho)} \prod_{i=1}^{l} \frac{\sin(i \pi \rho)}{\sin(\pi \rho)} 
\prod_{i=0}^{l-1}
\frac{\sin(\pi(1-2j_1\rho+i\rho))\sin(\pi(-2j_2\rho+i\rho))}{\sin(\pi(1-2j_1
\rho-2j_2
\rho+(l-1+i)\rho))}\nonumber \\
\times \prod_{i=0}^{s-l-1} \frac{\sin(\pi(-2j_3\rho+i \rho))
\sin(\pi(1+2j_1\rho+2j_3\rho+2j_2\rho-2\rho(s-1)+i \rho))}
{\sin(\pi(1+2j_1\rho+2j_2\rho-2\rho(s-1)+(s-l-1+i)\rho))}.
\end{eqnarray}

Following a related computation in 
\cite{dot2}, let us write
\begin{eqnarray}
\label{42}
z^{2j_1j_2\rho} (1-z)^{2j_2j_3\rho} {}_2F_1\left [ \left .\begin{array}{c}
-s+n,-j_3+m_3 \\ j_2-m_2-s+n+1 \end{array} \right | z \right ]
J_{n}^l(z) = \lambda_l(\rho)
N_{n}^l f_{n}^l(z)
z^{\gamma_n^l},
\end{eqnarray}
where $f_{n}^l(z)$ are regular functions of $z$ with $f_{n}^l(0)=1$,
so that the 
four-point function can be rewritten as
\begin{eqnarray}
\label{60}
{A}_4^{w=0}\left[
\begin{matrix}
j_1,j_2,j_3,j_4 \cr  j_1, m_2, m_3,m_4 \cr \end{matrix}\right] =
\Gamma(-s)  
\sum_{l=0}^s\sum_{n,\overline n =0}
^{s} X_{l} \left\{ z^{\gamma_n^l}
  B_n N_n^l f^l_n(z)\times c.c. \right\} .
\end{eqnarray}

The expressions for $N_{n}^l$ and $\gamma_n^l$ differ, depending on whether
$(a): l+n\le s$, or  $(b):l+n > s$.

Case $(a)$: In this case, we have
\begin{eqnarray}
\label{43}
\gamma_n^l = 2j_1j_2\rho-2lj_1\rho-2lj_2\rho+\rho l(l-1)\, ,
\end{eqnarray}
 and since it does not depend on $n$, we will denote it simply by
$\gamma_l$.

The normalization constant
$N_{n}^l$
 is obtained omitting the overall $z$-dependence in (\ref{27}) and afterwards
 taking the limit 
$z \rightarrow 0$, namely,
\begin{eqnarray}
\label{44}
N_{n}^l &=& \frac{1}{l!(s-l)!} \, \int_{(0,1)^{l}} \prod_{i=1}^l
du_{i} 
\prod_{q=1}^l u_q^{-2j_1\rho}
(1-u_q)^{-2j_2\rho-1}
\prod_{q<p}^l |u_q-u_p|^{2\rho} \nonumber \\
&\times& \int_{(1,\infty)^{s-l}} \prod_{i=1}^{s-l} dy_i \prod_{i=1}^{s-l}
y_i^{-2j_1\rho-2j_2\rho-1+2l\rho}
(y_i-1)^{-2j_3\rho-1}
\prod_{i<j}^{s-l} |y_i-y_j|^{2\rho}\alpha^{s-l}_{n}(y_1,\dots,y_{s-l}),
\end{eqnarray}
where we have used the identity
\begin{equation}
 \alpha^s_{n}(y_1,\dots,y_{s-l},0,\dots,0) = \alpha^{s-l}_{n}(y_1,\dots,
y_{s-l}). \label{228b}
\end{equation}

Let us denote
the Selberg integrals in the first line\footnote{Recall that we are using the 
notation
  introduced in 
\cite{in}.}  as
$S_l(-2j_1\rho+1,-2j_2\rho,\rho)$. The remaining integral
can be computed using Aomoto's formula.
Indeed, changing variables $y_i \rightarrow 1/y_i$ and
using the conservation laws (\ref{cj})-(\ref{consl3}), we can rewrite
the last line in (\ref{44}) as
\begin{eqnarray}
 A^{s-l-n}_{s-l}(-2j_4\rho,-2j_3\rho,\rho)
= \int_{(0,1)^{s-l}} \prod_{i=1}^{s-l} dy_i
\prod_{i=1}^{s-l} 
y_i^{-2j_4\rho-1} (1-y_i)^{-2j_3\rho-1} \prod_{i<j}^{s-l}
|y_i-y_j|^{2\rho} \alpha^{s-l}_{s-l-n}(y_1,\dots,y_{s-l}), \nonumber
\end{eqnarray}
where we have used the identity
\begin{eqnarray}
\alpha^{s-l}_{n}\left(\frac{1}{y_1},\dots,\frac{1}{y_{s-l}}\right) =
\left[\prod_{i=1}^{s-l}
y_i^{-1} \right] \alpha^{s-l}_{s-l-n}(y_1,\dots,y_{s-l}). \nonumber
\end{eqnarray}

Therefore, the normalization constant may be written as
\begin{eqnarray}
\label{50}
N_{n}^l = \frac{1}{l!(s-l)!} \, S_l(-2j_1\rho+1,-2j_2\rho,\rho)
A_{s-l}^{s-l-n}
(-2j_4\rho,-2j_3\rho,\rho).
\end{eqnarray}

Case $(b)$: When $l+n>s$ we have
\begin{eqnarray}
\label{51}
\gamma_n^l = \gamma_l
+l+n-s,
\end{eqnarray}
and
\begin{eqnarray}
 N_{n}^l &=& \frac{1}{l!(s-l)!} \, \int_{(0,1)^{l}} \prod_{i=1}^{l} du_i
\prod_{q=1}^l u_q^{-2j_1\rho}
(1-u_q)^{-2j_2\rho-1} \prod_{q<p}^l |u_q-u_p|^{2\rho}
\alpha^{l}_{n+l-s}(u_1,\dots,u_l)  \nonumber \\
&& \times ~ 
\int_{(0,1)^{s-l}} \prod_{i=1}^{s-l} dy_i \prod_{i=1}^{s-l} y_i^{-2j_4\rho-1}
(1-y_i)^{-2j_3\rho-1}\prod_{i<j}^{s-l}
|y_i-y_j|^{2\rho}, \nonumber
\end{eqnarray}
where we have used
\begin{eqnarray}
\lim\limits_{z\rightarrow 0} z^{s-l-n}
\alpha^s_{n}(y_1,\dots,y_{s-l},zu_1,\dots,zu_l)
= \left[ \prod_{i=1}^{s-l} y_i \right]
\alpha^{l}_{n+l-s}(u_1,\dots,u_l), \label{230b}
\end{eqnarray}
and therefore we get
\begin{eqnarray}
\label{500}
 N_{n}^l = \frac{1}{l!(s-l)!} \, 
A_{l}^{l+n-s}(-2j_1\rho+1,-2j_2\rho,\rho)S_{s-l}(-2j_4\rho,-2j_3\rho,\rho).
\end{eqnarray}

Notice that the values of $\gamma_n^l$ given by (\ref{51})
are always greater than
those in
(\ref{43}) and thus they do not
contribute to the lowest order in the factorization limit.

Using the identities (see \cite{dotfat} and \cite{in}):
\begin{eqnarray}
&& {\cal S}_l(a,b,\rho) = \frac{1}{l!} \, \prod_{i=1}^{l} \frac{\sin(i
  \pi \rho)}
{\sin(\pi\rho)} \prod_{i=0}^{l-1}
  \frac{\sin(\pi(a-1+i\rho))\sin(\pi(b-1+i\rho))}
{\sin(\pi(a+b-2+(l-1+i)\rho))} S_l(a,b,\rho)^2,\nonumber
\end{eqnarray}
\begin{eqnarray}
\label{64}
 {\cal A}_l^{n,\overline{n}}(a,b,\rho) = \frac{1}{l!} \,
 \prod_{i=1}^{l}
\frac{\sin(i \pi \rho)}{\sin(\pi\rho)} \prod_{i=0}^{l-1}
 \frac{\sin(\pi(a-1+i\rho))
\sin(\pi(b-1+i\rho))}{\sin(\pi(a+b-2+(l-1+i)\rho))} A_l^n(a,b,\rho)
A_l^{\overline{n}}(a,b,\rho),
\end{eqnarray}
and replacing (\ref{bn}), (\ref{59}), (\ref{43}), (\ref{50}),
 (\ref{51}) and (\ref{500})
 into (\ref{60}), it follows that the four-point amplitude may be
 rewritten in the 
following useful form:
\begin{eqnarray}
&& A_4^{w=0}\left[
\begin{matrix}
j_1,j_2,j_3,j_4 \cr  j_1, m_2, m_3,m_4 \cr \end{matrix}\right] \nonumber \\
&&~~~~~~
 =\Gamma(-s) \sum_{l=0}^s |z|^{2\gamma_l} {s \choose l}\left [\sum_{n,\overline n=0}^
{s-l}
|B_{s-l-n}|^2
{\cal S}_l(-2j_1\rho+1,-2j_2\rho,\rho)  {\cal
 A}_{s-l}^{n,\overline{n}}
(-2j_4\rho,-2j_3\rho,\rho) |f^l_{s-l-n}(z)|^2 \right.
\nonumber \\
&&~~~~~~~~\left.+ \sum_{n,\overline n=1}^
{l}
 z^{n}\bar z^{\bar n} |B_{s-l+n}|^2
{\cal S}_{s-l}(-2j_4\rho,-2j_3\rho,\rho)  {\cal
 A}_{l}^{n,\overline{n}}
(-2j_1\rho+1,-2j_2\rho,\rho) |f^l_{s-l+n}(z)|^2\right ].
\label{4pt}
\end{eqnarray}

Recall that this expression holds for integer numbers of screening
charges
 and it involves one highest-weight
operator. It is possible to relax the highest-weight restriction using the
Campbell-Backer-Hausdorff identity and proceeding as was done
for the easier case  of  the three-point functions in
\cite{becker2,in}. We shall not perform
this tedious calculation here but we show later that the
leading terms in the $z, \overline z \rightarrow 0$ limit of (\ref{4pt})
 can be identified with products of
three-point functions from where it is straightforward to see that
relaxing the highest-weight condition turns the Selberg
integrals into  combinations of Aomoto integrals.

In order to have a closed form for the conformal
blocks\footnote{Although they are 
closely related,
  the functions $f_n^l(z)$ should be distinguished from $f_n(x)$
 introduced in (\ref{4ptt}).} $f_n^l(z)$
 one must solve the multiple integrals in (\ref{17}) and this is a
difficult task.\footnote{Actually,  (\ref{17}) is obtained when the
amplitude involves one highest-weight state. The most general
expression for the conformal blocks when four generic states are
considered can be reconstructed replacing 
$\alpha^s_{n}(y_1, \dots,y_s)$ 
by 
$s_{nr}(y_1, \dots , y_s)$
and the coefficients
$C_{nr}$ given in (\ref{sc}) by (\ref{pq}).
In the limit $z, \overline z \rightarrow 0$,
 the integrals will reduce to those computed in 
\cite{sergio}, but the manipulations
performed in this section with the 
elementary symmetric  polynomials   cannot be easily
generalized when arbitrary Schur
polynomials  are involved.}
Even in the simpler cases of Liouville theory or the $H_3^+$-WZNW
model no
explicit formula for the conformal blocks is
 known. However, albeit
 the existence of a closed expression is unlikely, we shall be able
to examine the  leading terms in the factorization limit and perform the analytic
  continuation of (\ref{4pt}) to
generic values of the spins.

\section{The factorization limit}

In this section we  study the leading terms in the
factorization limit of the four-point function, $i.e.$ we 
retain only the leading terms in the $z, \overline z\rightarrow 0$
limit of (\ref{4pt}) and examine the following expression:
\begin{eqnarray}
&& {\mathbb A}_4^{w=0}\left[
\begin{matrix}
j_1,j_2,j_3,j_4 \cr  j_1, m_2, m_3,m_4 \cr \end{matrix}\right] \nonumber\\
&&~~~~~~~~~ ~~\equiv \Gamma(-s) \sum_{l=0}^s\sum_{n,\overline n=0}^
{s-l}
 |z|^{2\gamma_l} {s \choose l} |B_{s-l-n}|^2
{\cal S}_l(-2j_1\rho+1,-2j_2\rho,\rho)  {\cal
 A}_{s-l}^{n,\overline{n}}
(-2j_4\rho,-2j_3\rho,\rho).
\label{a4}
\end{eqnarray}

\subsection{Identification of the intermediate channels}

The leading powers of $z$ in the factorization of the amplitude of four
unflowed states
are expected to 
be of the form
$\hat{\Delta}_{j,m,w}-\Delta_{j_1}-\Delta_{j_2}$.
In general there are 
various
choices of quantum numbers for which this combination equals $\gamma_l$,
and then the
 intermediate channels cannot be unambiguously determined
from this
equality. No ambiguities  arise 
 in the semiclassical regime where only unflowed 
operators are expected to appear, so that equating 
$\gamma_l=\Delta_{j}-\Delta_{j_1}-\Delta_{j_2}$ one can read the possible 
values of the spin of the intermediate states. They are given by $j\equiv j_0=
-1-j_1-j_2+l$ and $j=-1-j_0$, in agreement with \cite{dot2}.

Consistently with this identification, 
let us now show that the  sums over $n$ and $\overline n$ in
(\ref{a4})
can be rewritten as
 products of two $w=0$ three-point functions divided by the two-point function
of the unflowed intermediate state. 

Recall the expression for the three-point functions given by
Eq.~(3.42) in \cite{in}.
Changing labels, this equation may be rewritten as
\begin{eqnarray}
\label{69}
&&A_3^{w=0}\left[
\begin{matrix}
j_4,j_3,-1-j \cr  m_4, m_3, m \cr \end{matrix}\right] =
 \Gamma(-s+l)\left[\frac{\Gamma(1+j_4-m_4)}{\Gamma(-s+l+1+j_4+j_3-m_4-m_3)}
\times c.c.\right]
\nonumber \\
&&~~~~~\times
\sum_{n,\overline{n}=0}^{s-l}(-1)^{n+\overline{n}}
\left[\frac{\Gamma(-s+l+1+j_4+j_3-m_4-m_3+n)}{\Gamma(1-s+l+j_4-m_4+n)}
\times c.c.\right]
\mathcal{A}_{s-l}^{n,\overline{n}}(-2j_4\rho,-2j_3\rho,\rho),
\label{tpf}
\end{eqnarray}
where  $j_3+j_4-j=s-l$, $m=-m_3-m_4$
and ${\cal A}_{l}^{n,\overline n}(-2j_1\rho,-2j_2\rho,\rho)$ was
defined in 
(\ref{64}). 
The insertion
points of the operators in the three-point functions  are
taken at $(0,1,+\infty)$ and, as before, we omit the obvious
$\overline m$-dependence from the arguments for short.

Using the conservation laws for the original four-point function,
$i.e.~ j_2=s-1-j_1-j_3-j_4$,
$m_2=-j_1-m_3-m_4$, and recalling that
\begin{eqnarray}
{B}_{s-l-n} = \frac{\Gamma(1+j_4-m_4)}{\Gamma(l+n-s+1+j_4-m_4)}
\frac{\Gamma(s-j_3-j_4+m_3+m_4)}{\Gamma(s-j_3-j_4+m_3+m_4-l-n )}, \nonumber
\end{eqnarray}
the three-point function (\ref{tpf}) may be rewritten as
\begin{eqnarray}
 A_3^{w=0}\left[
\begin{matrix}
j_4,j_3,-1-j \cr  m_4, m_3, m \cr \end{matrix}\right] &=& \Gamma(-s+l) \,
\frac{\Gamma(1-s+j_3+j_4-m_3-m_4)}{\Gamma(-s+l+1+j_4+j_3-m_4-m_3)}
\nonumber \\
&\times & \frac{\Gamma(1-s+j_3+j_4-\overline{m}_3-\overline{m}_4)}
{\Gamma(-s+l+1+j_4+j_3-\overline{m}_4-\overline{m}_3)}
\sum_{n,\overline{n}=0}^{s-l}
|B_{s-l-n}|^2
\mathcal{A}_{s-l}^{n,\overline{n}}(-2j_4\rho,-2j_3\rho,\rho).~~~~~~\nonumber
\end{eqnarray}

Therefore, it follows from (\ref{a4}) that
\begin{eqnarray}
\label{76}
{\mathbb A}_4^{w=0}\left[
\begin{matrix}
j_1,j_2,j_3,j_4 \cr  j_1, m_2, m_3,m_4 \cr \end{matrix}\right] &=&
\Gamma(-s) 
\sum_{l=0}^s |z|^{2
\gamma_l} {s \choose l} {\cal S}_l(-2j_1
\rho+1,-2j_2\rho,\rho)  \nonumber \\
& \times& \frac{\Gamma(-j_2+m_2+l)\Gamma(-j_2+\overline{m}_2+l)}
{\Gamma(-s+l)\Gamma(-j_2+m_2)\Gamma(-j_2+\overline{m}_2)}A_3^{w=0}\left[
\begin{matrix}
j_4,j_3,-1-j \cr  m_4, m_3, m \cr \end{matrix}\right].
\end{eqnarray}

Using the following identity  proved in \cite{becker2}:
\begin{eqnarray}
 A_3^{w=0}\left[
\begin{matrix}
j_1,j_2,j \cr  j_1, m_2, -m \cr \end{matrix}\right] =
 \Gamma(-l)\,\frac{\Gamma(-j_2+m_2+l)\Gamma(-j_2+\overline{m}_2+l)}{\Gamma(-j_2+m_2)
\Gamma(-j_2+\overline{m}_2)} {\cal S}_l(-2j_1\rho+1,-2j_2\rho,\rho),
\nonumber
\end{eqnarray}
where $j_1+j_2+j+1=l$,  (\ref{76}) may be recast as 
\begin{eqnarray}
 {\mathbb A}_4^{w=0}\left[
\begin{matrix}
j_1,j_2,j_3,j_4 \cr  j_1, m_2, m_3,m_4 \cr \end{matrix}\right] 
& =& \frac{1}{ \Gamma(0)}
\sum_{l=0}^s |z|^{2\gamma_l}
A_3^{w=0}\left[
\begin{matrix}
j_1,j_2,j \cr  j_1, m_2, -m \cr \end{matrix}\right]A_3^{w=0}\left[
\begin{matrix}
j_4,j_3,-1-j \cr  m_4, m_3, m \cr \end{matrix}\right]\nonumber\\
& =&
\sum_{l=0}^s |z|^{2\gamma_l}
A_3^{w=0}\left[
\begin{matrix}
j_1,j_2,j \cr  j_1, m_2, -m \cr \end{matrix}\right]A_3^{w=0}\left[
\begin{matrix}
j_4,j_3,-1-j \cr  m_4, m_3, m \cr \end{matrix}\right]A_2^{w=0}\left[
\begin{matrix}
j,-1-j \cr  -m, m \cr \end{matrix}\right]^{-1} , \nonumber
\end{eqnarray}
where we have used
$ \Gamma(-l) l! = (-1)^l \Gamma(0)$ in the first line and 
the factor $\Gamma(0)$ has been interpreted as the $\delta^2(0)$
 from a two-point function  in the
second line.

At this point we can straightforwardly relax the highest-weight
 condition of the state at $z_1,\overline z_1=0$
using the Backer-Campbell-Hausdorff formula as in \cite{becker2} to finally get:
\begin{eqnarray}
\label{83b}
 {\mathbb A}_4^{w=0}\left[
\begin{matrix}
j_1,j_2,j_3,j_4 \cr  m_1, m_2, m_3,m_4 \cr \end{matrix}\right] 
 = \sum_{l=0}^s |z|^{2\gamma_l} A_3^{w=0}\left[
\begin{matrix}
j_1,j_2,j \cr  m_1, m_2, -m \cr \end{matrix}\right] A_3^{w=0}\left[
\begin{matrix}
j_3,j_4,-1-j \cr  m_3,m_4,m \cr \end{matrix}\right] A_2^{w=0}\left[
\begin{matrix}
j,-1-j \cr  -m, m \cr \end{matrix}\right]^{-1},
\end{eqnarray}
where $j=j_0\equiv -1-j_1-j_2+l$ (alternatively, $j=-1-j_0$) and $m=m_1+m_2=-m_3-m_4$.

Changing the index $l \rightarrow (s-l)$
in  (\ref{83b}) 
we get
another parametrization of the four-point function that will be important
when discussing 
its analytic continuation below, namely
\begin{eqnarray}
\label{84}
 {\mathbb A}_4^{w=0}\left[
\begin{matrix}
j_1,j_2,j_3,j_4 \cr  m_1, m_2, m_3,m_4 \cr \end{matrix}\right] 
= \sum_{l=0}^s |z|^{2\gamma'_l}A_3^{w=0}\left[
\begin{matrix}
j_1,j_2,-1-j' \cr  m_1, m_2, -m \cr \end{matrix}\right]A_3^{w=0}\left[
\begin{matrix}
j_4,j_3,j' \cr  m_4, m_3, m \cr \end{matrix}\right]A_2^{w=0}\left[
\begin{matrix}
-1-j',j' \cr  -m,m \cr \end{matrix}\right]^{-1},
\end{eqnarray}
where $\gamma_l'$ equals $\gamma_l$  with the replacement 
$j \rightarrow j'=-1-j_3-j_4+l$.

 Eq.~(\ref{83b}) 
expresses the 
content of the factorization limit of the four-point functions 
obtained in the
 Coulomb gas 
approach in the semiclassical limit.
However, this expression  was
 deduced without making 
any assumption on the values of $k$, except for the identification of 
the intermediate channels with unflowed operators. It is surprising that 
all the terms in (\ref{a4}) 
can be identified as contributions of $w=0$
intermediate states
because
it was shown in \cite{satoh} 
that the
OPE of unflowed states (when defined as in \cite{tesch3}) 
contains contributions
from operators 
outside the physical spectrum of the $AdS_3$-WZNW
model and it was argued in \cite{wc} that the spectral flow symmetry of
 the model requires to additionally consider
$w$-violating structure constants.
In section 3.3 we elaborate on these issues.

\subsection{Analytic continuation}

In order to perform the analytic continuation of ${\mathbb A}_4^{w=0}$
for generic external states, notice that
the integer nature of the number of screening charges
is encoded both in the upper limit of the sum in (\ref{83b}) and in
the fact that this 
expression is actually a discrete sum: recall that the first three-point
function in this equation
involves $l$ screening operators while the second one
 involves the remaining $s-l$ ones. In order to obtain an expression for
 generic 
unflowed external states we will 
identify the terms
in the sum over $l$
with the residues of a meromorphic function extending the summands
sequence. This will 
allow us
to rewrite the four-point correlator
as a complex integral where the integer nature of the
 number of screening operators will be strictly restricted to the
 choice of the 
integration
contour. For a suitable set of the kinematical parameters
this contour can be fixed and generic amplitudes can be obtained. This strategy
to perform the analytic continuation of (\ref{83b}) 
to generic spin values
of the external states in the semiclassical limit
is inspired by \cite{tesch2}. In the next subsection we discuss the
validity of the result for arbitrary values of the affine level $k$.

In order to trade the sum in (\ref{83b}) for an integral, let us first notice that
it can be freely extended to $\infty$ (see (\ref{a4})). Furthermore, given
that the 
two-point function in the denominator of (\ref{83b})
diverges as $\Gamma(0)$, we can
use the following identity, which is valid for any sequence $K(l)$:
\begin{eqnarray}
\label{94}
 \frac{1}{\Gamma(0)} \, \sum_{l=0}^{\infty} K(l) = \frac{1}{2\pi i}
\oint_{\cal C} {\cal K}(x) \, dx ,
\end{eqnarray}
where ${\cal K}(x)$ is a meromorphic continuation of $K(l)$ having
simple poles at 
$x=0,1,2,\dots,\infty$, with $K(x)$ behaving as $\Gamma(-x)$ near 
them\footnote{Eq.~(\ref{94}) is a suitable form of
  the classical
N\"orlund-Rice theorem for infinite sums, which states that
\begin{eqnarray}
\label{88}
&& \sum_{l=0}^{\infty} \frac{(-1)^{l}}{l!} H(l) =
\frac{1}{2\pi i} \oint\limits_{\mathcal{C}} \Gamma(-x) {\cal H}(x) \, dx\, ,
\end{eqnarray}
for any meromorphic continuation ${\cal H}(x)$ of $H(l)$ having no
poles in the 
positive integer numbers. Eq.~(\ref{94}) is obtained from (\ref{88})
after using 
the formal expression $ l! = \Gamma(1+l) =
{(-1)^l \Gamma(0)}/{\Gamma(-l)}$.}.
 The contour ${\cal C}$
is understood to
enclose only these poles and neither of the other poles that ${\cal K}(x)$ could have.

Our first task is to find a proper analytic continuation for
the sequence of 
summands in (\ref{83b}).
To this aim, recall that
it was proved in \cite{in} that the Coulomb gas
representation of the
three-point functions $A_3^{w=0}\left[
\begin{matrix}
j_1,j_2,j_3 \cr  m_1,m_2,m_3 \cr \end{matrix}\right]$ admits such
 analytic continuation in the number of screening operators
leading to the following exact expression \cite{tesch1, satoh}:
\begin{eqnarray}
\mathcal{A}_3^{w=0}\left[
\begin{matrix}
j_1,j_2,j_3 \cr  m_1,m_2,m_3 \cr \end{matrix}\right] =
\delta^2\left(m_1+m_2+m_3 \right) 
D(-1-j_1,-1-j_2,-1-j_3) W\left[
\begin{matrix}
j_1,j_2,j_3 \cr  m_1, m_2, m_3 \cr \end{matrix}\right], \nonumber
\end{eqnarray}
where we have introduced the $\delta^2\left(m_1+m_2+m_3 \right)$ in
order to reinforce 
the conservation law implicit in the free field computation,
$D(j_1,j_2,j_3)$ is the 
structure constant given by
\begin{eqnarray}
\label{97}
D(j_1,j_2,j_3) =
\frac{G(1+j_1+j_2+j_3)G(j_1+j_2-j_3)G(j_2+j_3-j_1)G(j_3+j_1-j_2)}
{\nu^{-j_1-j_2-j_3-1} G_0 G(1+2j_1) G(1+2j_2)  G(1+2j_3)},
\end{eqnarray}
with
\begin{eqnarray}
G(j)=(k-2)^{\frac{j(1-j-k)}{2(k-2)}} \, \Gamma_2(-j |1,k-2 ) \,
\Gamma_2(k-1+j |1,k-2 ), \nonumber
\end{eqnarray}
$\Gamma_2(x|1,w)$ being the Barnes double Gamma function,
$G_0 = -2 \pi^2 \gamma\left(1-\rho\right) G(-1)$
and
\begin{eqnarray}
 W\left[
\begin{matrix}
j_1,j_2,j_3 \cr  m_1, m_2, m_3 \cr \end{matrix}\right]
 = \int d^2x_1\, d^2x_2\, x_1^{j_1+m_1}
\overline{x}_1^{j_1+\overline{m}_1} x_2^{j_2+m_2}
\overline{x}_2^{j_2+\overline{m}_2}
 |1-x_1|^{-2j_{13}-2}
|1-x_2|^{-2j_{23}-2}
|x_1-x_2|^{-2j_{12}-2}.\nonumber
\end{eqnarray}

This function  was computed in \cite{hh} and it was shown in
\cite{satoh} that it
 reduces to 
\begin{eqnarray}
 W_1\left[
\begin{matrix}
j_1,j_2,j_3 \cr  m_1, m_2, m_3 \cr \end{matrix}\right]&=&
 (-1)^{\overline{m}_3-m_3+
\overline{q}_1}\pi^2\frac{
 \gamma(-1-j_1-j_2-j_3)\gamma(1+2j_1)}{\gamma(1+j_{12})
\gamma(1+j_{13})}
 \frac{\Gamma(1+j_2-m_2)}{\Gamma(-j_2+\overline{m}_2)}
\frac{\Gamma(1+j_3-m_3)}{\Gamma(-j_3+\overline{m}_3)}  \nonumber \\
& \times &
\left\{ \frac{\Gamma(1+j_3+m_3)}{\Gamma(1+j_3+m_3-q_1)}
{}_3F_2\left [ \left .\begin{array}{c}
-q_1,\,-j_{12},\,1+j_{23} \\ -2j_1,\,1+j_3+m_3-q_1 \end{array} \right | 1 \right ]
\times \mbox{c.c.}\right\} ,
\label{w1}
\end{eqnarray}
for $m_1=j_1-q_1$ and $\overline{m}_1=j_1-\overline{q}_1$ 
with $q_1,\overline{q}_1=0,1,2,\dots$.

In order to analyze the analytic structure of the summand in
(\ref{83b}), it is
useful to parametrize both three-point functions in a similar way by
using the following 
identity:
\begin{eqnarray}
{\cal A}_3^{w=0}\left[
\begin{matrix}
j_3,j_4,-1-j \cr  m_3,m_4,m \cr \end{matrix}\right] {\cal A}_2^{w=0}\left[
\begin{matrix}
j,-1-j \cr  -m, m \cr \end{matrix}\right]^{-1}={\cal A}_3^{w=0}\left[
\begin{matrix}
j_3,j_4,j \cr  m_3,m_4,m \cr \end{matrix}\right] {\cal A}_2^{w=0}\left[
\begin{matrix}
j,j \cr  -m, m \cr \end{matrix}\right]^{-1}, \nonumber
\end{eqnarray}
which follows from (2.6)-(2.7) and (3.5) in \cite{satoh}, so that, up
to an irrelevant 
factor, we can rewrite (\ref{83b})
as the following integral:
\begin{eqnarray}
\label{1032}
&&  {\mathbb A}_4^{w=0}\left[
\begin{matrix}
j_1,j_2,j_3,j_4 \cr  m_1, m_2, m_3,m_4 \cr \end{matrix}\right]  \nonumber \\
&& ~~~~~~~~ = \oint\limits_{\mathcal{C}}
 |z|^{2(\Delta_{j}-\Delta_{j_1}-
\Delta_{j_2})} {\cal A}_3^{w=0}\left[
\begin{matrix}
j_1,j_2,j \cr  m_1,m_2,-m \cr \end{matrix}\right] {\cal A}_3^{w=0}\left[
\begin{matrix}
j_3,j_4,j \cr  m_3,m_4,m \cr \end{matrix}\right] {\cal A}_2^{w=0}\left[
\begin{matrix}
j,j \cr  -m, m \cr \end{matrix}\right]^{-1} dj,
\end{eqnarray}
where ${\cal C}$ encloses the poles at $j=-1-j_1-j_2+{\mathbb Z}_{\ge 0}$.

Some care must be taken when applying (\ref{94}) since it is valid for
 a meromorphic extension behaving as a gamma function near the integer
 poles. 
The three-point function
 ${\cal A}_3^{w=0}\left[
\begin{matrix}
j_1,j_2,j \cr  m_1,m_2,-m \cr \end{matrix}\right]$ exhibits this
behaviour near 
the poles
at $j=-1-j_1-j_2+{\mathbb Z}_{\ge 0}$ in the factor
$\gamma(-1-j_1-j_2-j)$ in (\ref{w1}). The structure constants
 have no such poles.

The fact that $s$ is an integer number plays no role in (\ref{1032}) and, in that sense, this expression can be thought to be valid even for $j_1+j_2+j_3+j_4+1 \notin \mathbb{N}_0$. Recall that the three-point functions involve fields with  generic kinematical configurations. However, it is important to notice that although the spins of the
external states
are no longer restricted, the integer nature of the number
of screening operators remains
encoded
in the prescription for the choice of the integration contour.
Indeed, it is necessary to specify ${\cal C}$ in order to have a well-defined
analytic continuation. This does not seem possible for arbitrary
configurations \cite{tesch2}, but in the semiclassical limit, one can freely set 
${\cal C}={\cal P}=-1/2 +i{\mathbb R}$ 
restricting 
the quantum numbers of the
external states as follows:
\begin{eqnarray}
\left\{
\begin{array}{ll}
|{\mbox{Re}}(j_1+j_2+1)|<\frac{1}{2}, \qquad & |{\mbox{Re}}(j_3+j_4+1)|<\frac{1}{2}, \\
|{\mbox{Re}}(j_1-j_2)|<\frac{1}{2}, \qquad & |{\mbox{Re}}(j_3-j_4)|<\frac{1}{2},
\label{condic}
\end{array}
\right.
\end{eqnarray}
\begin{eqnarray}
\left\{
\begin{array}{ll}
\max\{m_1+m_2,\overline{m}_1+\overline{m}_2\}>-\frac{1}{2}, \\
\min\{m_1+m_2,\overline{m}_1+\overline{m}_2\}<\frac{1}{2}.
\label{condic2}
\end{array}
\right.
\end{eqnarray}

Indeed, taking the $k\rightarrow \infty$ limit, the poles 
of the first three-point function in (\ref{1032}) are located at
\begin{eqnarray}
\left\{
\begin{array}{l}
j=-1-j_1-j_2+\mathbb{Z}_{\ge 0} , \\
j=j_2-j_1+\mathbb{Z}_{\ge 0} ,
\end{array}
\right.
\label{pol1bis}
\end{eqnarray}
\begin{eqnarray}
 \left\{
\begin{array}{l}
j=j_1+j_2-\mathbb{Z}_{\ge 0}  ,\\
j=j_1-j_2-1-\mathbb{Z}_{\ge 0} ,\end{array}
\right.
\label{pol1bis2}
\end{eqnarray}
\begin{eqnarray}
j=-\max\{m_1+m_2,\overline{m}_1+\overline{m}_2\}-\mathbb{Z}_{>0}\, ,
\nonumber
\end{eqnarray}
while those coming from the second three-point function are placed at
\begin{eqnarray}
\left\{
\begin{array}{l}
j=-1-j_3-j_4+\mathbb{Z}_{\ge 0} ,\\
j=j_4-j_3+\mathbb{Z}_{\ge 0} ,
\end{array}
\right.
\label{pol1biss}
\end{eqnarray}
\begin{eqnarray}
\left\{
\begin{array}{l}
j=j_3+j_4-\mathbb{Z}_{\ge 0} ,\\
j=j_3-j_4-1-\mathbb{Z}_{\ge 0} , \end{array}
\right.
\label{pol1biss2}
\end{eqnarray}
\begin{eqnarray}
j=\min\{m_1+m_2,\overline{m}_1+\overline{m}_2\}-\mathbb{Z}_{>0}\, .
\nonumber
\end{eqnarray}

Under (\ref{condic2}) the poles depending on $m_1,m_2$ and 
$\overline{m}_1,\overline{m}_2$ lie in the left half complex
$j-$plane, and this is 
also the case for the poles at (\ref{pol1bis2}) and (\ref{pol1biss2}). It
follows that 
closing the contour
at infinity to the right, the only poles encircled are at (\ref{pol1bis})
and 
(\ref{pol1biss}). By virtue of 
the parametrization 
 (\ref{84}) it is easy to see that the contributions from the residues
in both families of poles are identical. Finally, let us notice that
both series of 
poles in (\ref{pol1bis}) are
related by the reflection $j_2 \leftrightarrow (-1-j_2)$. It is proved
in Appendix A.2 
that the residues at
the second series of poles in (\ref{pol1bis}) vanish
if the state at $z_2,\overline{z}_2=1$ lies in a discrete series.

Summarizing, we have found, up to irrelevant factors, that the leading term in
the factorization of the four-point function is given, in the semiclassical limit, by
\begin{eqnarray}
&& {\mathbb A}_4^{w=0}\left[
\begin{matrix}
j_1,j_2,j_3,j_4 \cr  m_1, m_2, m_3,m_4 \cr \end{matrix}\right] 
 = \nonumber\\
&&~~~~~~~~~~~~~~~
\int_{\mathcal P}dj    ~
 |z|^{2(\Delta_{j}-\Delta_{j_1}-
\Delta_{j_2})} {\cal A}_3^{w=0}\left[
\begin{matrix}
j_1,j_2,j \cr  m_1,m_2,-m \cr \end{matrix}\right] {\cal A}_3^{w=0}\left[
\begin{matrix}
j_3,j_4,j \cr  m_3,m_4,m \cr \end{matrix}\right] {\cal A}_2^{w=0}\left[
\begin{matrix}
j,j \cr  -m, m \cr \end{matrix}\right]^{-1}. 
\label{final}
\end{eqnarray}
This
expression makes no reference at all to the integer nature of the number of
screening operators 
 and it is valid for external
states  restricted as in (\ref{condic}),
(\ref{condic2}). For other values of the kinematical parameters it must be
 defined by analytic continuation, as discussed in \cite{tesch1,tesch3}.

Eq.~(\ref{final}) agrees with the one obtained by
transforming 
the $x$-basis
integral formula (\ref{4ptt}) for the four-point function 
  to the $m$-basis\footnote{See
  \cite{wc} for this 
computation and \cite{mn} for an alternative representation
of the integral transform of (\ref{4ptt}) to the $m$-basis.}.
In the $H_3^+$-WZNW model,
(\ref{4ptt}) was obtained in \cite{tesch2} in
 the mini-superspace
limit, which describes a semiclassical region of the full theory, and it was
postulated to be valid for generic values of $k$
from the OPE of normalizable states and the factorization ansatz in
\cite{tesch1, tesch3}.
Here, we have deduced it also in the large-$k$ limit using the Coulomb
gas method. 
If $w$-conserving amplitudes in
the $AdS_3$-WZNW models are related by
 analytic continuation to correlators of the $H_3^+$-WZNW model,
 as it is widely believed, a similar postulate
 would allow to 
extend the validity of (\ref{final}) beyond the semiclassical
regime. Such conjecture 
however is more subtle in the Lorentzian model than in its Euclidean
counterpart due to the spectral flow representations. 
In the following subsection we discuss this issue.

\subsection{Factorization into spectral flow violating three-point functions}

 The OPE of normalizable states
determining the 
four-point
functions (\ref{4ptt}) in the $H_3^+$-WZNW model would give an
incorrect zero answer 
if used to compute
$w$-violating three-point functions in the $AdS_3$-WZNW model. Actually,
consistency with the spectral
flow selection rules leads  to the
following OPE for spectral flow images of primary fields in the 
$AdS_3$-WZNW model \cite{wc}:
\ber
\Phi_{m_1,\overline m_1}^{j_1,w_1}(z_1,\overline z_1)
\Phi_{m_2,\overline m_2}^{j_2,w_2}(z_2,\overline z_2)=\sum_{w=0,\pm 1}
\int_{\cal P} {\cal A}^{w}_3\left[
\begin{matrix}
j_1\,,j_2,\,-1-j_3\cr  m_1, m_2, -m_3\cr\end{matrix}\right]
z_{12}^{-\hat\Delta_{12}}
\overline
z_{12}^{-\overline{\hat\Delta}_{12}}\Phi_{m_3,\overline m_3}
^{j_3,w_3}(z_2, \overline z_2)\,dj_3+\cdots\, , \label{ope}
\eer
where $w=w_3-w_1-w_2$, $z_{12}=z_1-z_2$,
$\hat\Delta_{12}=\hat\Delta_{j_1,m_1,w_1}+\hat\Delta_{j_2,m_2,w_2}-
\hat\Delta_{j_3,m_3,w_3}$ and
\begin{eqnarray}
{\cal A}_3^{w=\pm 1}\left[
\begin{matrix}
j_1,j_2,j_3 \cr  m_1,m_2,m_3 \cr \end{matrix}\right] =
\frac{\delta^2\left(m_1+m_2+m_3
\mp k/2 \right)}{\gamma(j_1+j_2+j_3+3-k/2)} {\widetilde
  D}(-1-j_1,-1-j_2,-1-j_3) 
{\widetilde W}\left[
\begin{matrix}
j_1,j_2,j_3 \cr  \mp m_1, \mp m_2,
\mp m_3 \cr \end{matrix}\right]
\nonumber
\end{eqnarray}
with
\begin{eqnarray}
\widetilde{D}(j_1,j_2,j_3) \sim B(j_1) D\left (-\frac k2-j_1,j_2,j_3 \right ),
\nonumber
\end{eqnarray}
up to $k$-dependent, $j$-independent factors and
\begin{eqnarray}
  \widetilde{W}\left[\begin{matrix}
j_1\,,\,j_2\,,\,j_3\cr  m_1, m_2, m_3\cr\end{matrix}\right]=
\frac{\Gamma(1+j_1+m_1)}
{\Gamma(-j_1-\overline{m}_1)}
\frac{\Gamma(1+j_2+\overline m_2)}{\Gamma(-j_2-m_2)}
\frac{\Gamma(1+j_3+\overline{m}_3)} {\Gamma(-j_3-m_3)}. \nonumber
\end{eqnarray}

Although it is necessary to further truncate this OPE in order to
avoid inconsistencies with the spectral flow symmetry, the physical
mechanism determining the decoupling not being yet completely
understood, 
several successful checks have been 
performed on the fusion rules obtained
from (\ref{ope}). In particular, they reproduce the classical tensor
product of representations of $SL(2,{\mathbb R})$ in the
$k\rightarrow\infty$ limit and moreover, for generic $k>2$
 they establish the closure of the operator
algebra on the Hilbert space of the $AdS_3$-WZNW model determined in 
\cite{mo1}. 

The factorization ansatz based on this OPE
would give an expression for the $w$-conserving
four-point correlation functions 
involving both spectral flowed and unflowed intermediate states. 
This conclusion also follows from the results in \cite{mo3}, where
$w=1$ long strings were found in the $s$-channel factorization of the
four-point functions of $w=0$ short strings starting from
 (\ref{4ptt}), rewriting the integrand and moving the integration contour.
However, these observations pose  an apparent contradiction with (\ref{a4}).

To understand this issue,
recall that we have performed the Coulomb gas computation of the expectation 
value of
four unflowed vertices without any insertion of
 spectral flow operators, namely, we have taken $N_+=N_-=0$ in
 (\ref{very}). 
Therefore we should not expect to be able to
recognize $w$-violating three-point functions
in the factorization limit
since $w\ne 0$ amplitudes require insertions of 
vertices $\eta^\pm$. 
However, the full final result
for the $w$-conserving four-point function must be the same,
independently of the (even) number of these
insertions because
they simply act as picture changing operators. This suggests  either that
there are no $w$-violating channels or that both
channels  give equivalent expansions. These two possibilities
also follow 
if correlation
functions in the $AdS_3$-WZNW model are to be obtained by analytic
continuation from those in the  $H_3^+$-WZNW model,
because the spectral flow fields do not belong to
the spectrum of the Euclidean theory. The results in 
\cite{mo3, wc} force the conclusion
that both channels give equivalent contributions. However, 
we should not expect to be able to verify this equivalence in general
just by looking at the leading terms in the
factorization limit. Rather, 
since the spectral flow operation maps primaries into non-primaries,
a general proof of this statement would require
making explicit the higher order terms in (\ref{ope}) and possibly
some contour manipulations.

Despite these general arguments, in the remaining of this section
we  use the Coulomb gas approach to illustrate   in a particular example
the assertion that products of $w$-preserving and
 violating three-point functions give the same contributions to the 
$w$-conserving four-point functions.

Let us start by evaluating the following amplitude
\begin{eqnarray}
&& A_4^{w=0}\left[
\begin{matrix}
 j_1,j_2, j_3,j_4 \cr  j_1, m_2, - j_3,m_4 \cr 
w_1,w_2,w_3,w_4\cr\end{matrix}\right] \nonumber\\
&&~~~~~~~~~~~= \Gamma(-s) \left\langle
V^{ j_1,w_1}_{
 m_1= {\overline{m}}_1=  j_1}(0)
V^{j_2,w_2}_{m_2,\overline{m}_2}(z)
V^{ j_3,w_3}_{  m_3={\overline{m}}_3=- j_3}(1)
V^{j_4,w_4}_{m_4,\overline{m}_4}(+\infty)
\eta^-(\zeta^-) \eta^+(\zeta^+) \mathcal{Q}^s\right\rangle.
\label{corrfun}
\end{eqnarray}
The insertion of the spectral flow operators will be explained later. 

After performing the corresponding field contractions we get
\begin{eqnarray}
 A_4^{w=0}\left[
\begin{matrix}
 j_1,j_2, j_3,j_4 \cr  j_1, m_2, - j_3,m_4 \cr 
w_1,w_2,w_3,w_4\cr\end{matrix}\right] 
&=& \frac{\Gamma(-s)}{\pi^2\Gamma(0)^2} \left[ 
(\zeta^--z)^{-(j_2-m_2)} (1-\zeta^-)^{-2 j_3}
(\zeta^+-z)^{-(j_2+m_2)}
(\zeta^--\zeta^+)^{-(k-1)} \right.\nonumber\\
&&\left. \times ~ (\zeta^+)^{-2 j_1}  z^{2 j_1j_2\rho -
\frac k2 w_1 w_2 - w_1m_2-w_2 j_1} (1-z)^{2j_2 j_3\rho -
\frac k2 w_2 w_3 + w_2 j_3-w_3m_2}  
\times {\mbox{c.c}}
\right] \nonumber 
\end{eqnarray}
\begin{eqnarray}
&&~~~~~~~~~~\times ~\int \prod_{i=1}^s d^2 y_i |y_i|^{-4 j_1\rho} 
|z-y_i|^{-4j_2\rho}
|1-y_i|^{-4 j_3\rho} |\zeta^+-y_i|^{4} \prod_{i<j}^s
|y_i-y_j|^{4\rho} \left[\frac{1}{\mathcal{P} }\,
\partial_1\dots\partial_s(\Lambda\mathcal{P})\times {\mbox{c.c}}\right] 
, \nonumber
\end{eqnarray}
where we have defined
\begin{eqnarray}
\mathcal{P} = \prod_{i=1}^{s}
 (z-y_i)^{-(j_2-m_2)} (1-y_i)^{-2 j_3}
(\zeta^+-y_i)^{-(k-2)} \prod_{i<j} (y_i-y_j) \qquad
{\rm and} \qquad
\Lambda = \prod_{i=1}^{s} \frac{\zeta^--y_i}{\zeta^+-y_i}.\nonumber
\end{eqnarray}
It was shown in \cite{in} that this expression reproduces the one
obtained using the prescription introduced in \cite{zz} to compute
correlators involving spectral flowed operators. Therefore, the dependence on
$\zeta^{\pm}$ and $\overline{\zeta}^{\pm}$ cancels and we can freely
take\footnote{Notice that this can be done only because there is 
a highest-weight state inserted at $z_1,\overline{z}_1=0$ and a lowest-weight
state at
 $z_3,\overline{z}_3=1$.}
$\zeta^-=\overline{\zeta}^-=0$ and $\zeta^+=\overline{\zeta}^+=1$,
obtaining
\begin{eqnarray}
&& A_4^{w=0}\left[
\begin{matrix}
 j_1,j_2, j_3,j_4 \cr  j_1, m_2, - j_3,m_4 \cr 
w_1,w_2,w_3,w_4\cr\end{matrix}\right] \nonumber\\
&&~~~~~~~~~~~~
= \frac{\Gamma(-s)}{\pi^2\Gamma(0)^2} \left[ z^{2 j_1j_2\rho - \frac k2 w_1 w_2 -
w_1m_2-w_2 j_1-j_2+m_2} (1-z)^{2j_2j_3\rho - \frac k2 w_2 w_3 +
 w_2 j_3-
w_3m_2-j_2-m_2} \times {\mbox{c.c}} \right] \nonumber \\
&&~~~~~~~~~~~~
 \times \int \prod_{i=1}^s d^2 y_i |y_i|^{-4j_1\rho+2} |z-y_i|^{-4j_2\rho}
|1-y_i|^{-4 j_3\rho+2} \prod_{i<j}^s |y_i-y_j|^{4\rho} \left[\frac{1}
{\mathcal{P}' }\,\partial_1\dots\partial_s\mathcal{P}'\times {\mbox{c.c}}
\right]\nonumber
\end{eqnarray}
where
\begin{eqnarray}
\mathcal{P}' = \prod_{i=1}^{s} y_i(z-y_i)^{-(j_2-m_2)}
(1-y_i)^{-2 j_3-k+1}
\prod_{i<j} (y_i-y_j). \nonumber
\end{eqnarray}
It is easy to check that
this expression equals, up to the factor
$(\pi\Gamma(0))^{-2}$,
the Coulomb integral realization of the amplitude
$
{A}_4^{w=0}\left[
\begin{matrix}
-1-\tilde j_1,j_2, -1-\tilde j_3,j_4 \cr  ~ -\tilde j_1,~ m_2, ~~~
\tilde j_3,~~ m_4 \cr
w_1-1,w_2,w_3+1,w_4\cr \end{matrix}\right],$ 
 where we have introduced
\begin{eqnarray}
\tilde j_1=-\frac k2 -j_1, \qquad \qquad \tilde j_3=-\frac k2 -j_3\, . 
\nonumber
\end{eqnarray}

Notice that the conservation laws for this correlation function, when
no spectral flow operators are inserted,
reproduce those of 
(\ref{corrfun}). Indeed, the spectral flow operators were inserted
in (\ref{corrfun}) to achieve this equality.

Using the reflection identity \cite{tesch3, hs, gk} and
$c^{\tilde{j}}_{\pm\tilde{j},\pm\tilde{j}} = \pi 
\Gamma(0)$,
we can write
\begin{eqnarray}
 V^{-1-\tilde{j},w}_{\pm\tilde{j},\pm\tilde{j}}=
 B(\tilde{j}) \, c^{\tilde{j}}_{\pm\tilde{j},\pm\tilde{j}}
V^{\tilde{j},w}_{\pm\tilde{j},\pm\tilde{j}} =
 \pi \Gamma(0)B(\tilde{j}) V^{\tilde{j},w}_{\pm\tilde{j},\pm\tilde{j}},
\nonumber
\end{eqnarray}
and then it is straightforward to show that
\begin{eqnarray}
&&A_4^{w=0}\left[
\begin{matrix}
 j_1,j_2, j_3,j_4 \cr  j_1, m_2, - j_3,m_4 \cr 
w_1,w_2,w_3,w_4\cr \end{matrix}\right] 
=
B(\tilde{j}_1) 
B(\tilde{j}_3) {A}_4^{w=0}\left[
\begin{matrix}
\tilde j_1,~~~j_2, ~~~~\tilde j_3,~~j_4 \cr   -\tilde j_1,~ ~m_2,~~ ~\tilde j_3,~~m_4
\cr w_1-1, w_2,  w_3+1, w_4\cr \end{matrix}
\right]\nonumber\\
&&~~~~~=z^{m_2+\frac k2 w_2}
\overline{z}^{\overline{m}_2+\frac k2 w_2} (1-z)^{-m_2-\frac k2 w_2}
(1-\overline{z})^{-\overline{m}_2-\frac k2 w_2}
B(\tilde{j}_1) 
B(\tilde{j}_3) {A}_4^{w=0}\left[
\begin{matrix}
\tilde j_1,j_2, \tilde j_3,j_4 \cr   -\tilde j_1, m_2, \tilde j_3,m_4
\cr w_1, w_2, w_3, w_4\cr \end{matrix}
\right].
\label{47}
\end{eqnarray}

This identity was assumed  in \cite{wc} as the starting
point of the proof 
that products of $w$-conserving
or $w$-violating three-point functions
 give the same contribution
when factorizing these four-point functions.
To be more explicit, let us show this statement in the particular case $w_i=0$,
$i=1,\cdots , 4$.

On the one hand, we have already found that (see Eq. (\ref{83b}) and the
discussion below)
\begin{eqnarray}
 {\mathbb A}_4^{w=0}\left[
\begin{matrix}
j_1,j_2,j_3,j_4 \cr  j_1, m_2, -j_3,m_4 \cr \end{matrix}\right] 
 = \sum_{l=0}^s |z|^{2(\Delta_{j}-\Delta_{j_1}-
\Delta_{j_2})} A_3^{w=0}\left[
\begin{matrix}
j_1,j_2,j \cr  j_1, m_2, -m \cr \end{matrix}\right] A_3^{w=0}\left[
\begin{matrix}
j_3,j_4,j \cr  -j_3,m_4,m \cr \end{matrix}\right] A_2^{w=0}\left[
\begin{matrix}
j,j \cr  -m, m \cr \end{matrix}\right]^{-1}
\label{serserr}
\end{eqnarray}
where $j=-1-j_1-j_2+l$, and we showed that this expression can be analytically
continued as
\begin{eqnarray}
{\mathbb A}_4^{w=0}\left[
\begin{matrix}
j_1,j_2,j_3,j_4 \cr  j_1, m_2, -j_3,m_4 \cr \end{matrix}\right] 
 = 
\int_{\mathcal P}dj    ~
 |z|^{2(\Delta_{j}-\Delta_{j_1}-
\Delta_{j_2})} {\cal A}_3^{w=0}\left[
\begin{matrix}
j_1,j_2,j \cr  j_1,m_2,-m \cr \end{matrix}\right] {\cal A}_3^{w=0}\left[
\begin{matrix}
j_3,j_4,j \cr  -j_3,j_4,m \cr \end{matrix}\right] {\cal A}_2^{w=0}\left[
\begin{matrix}
j,j \cr  -m, m \cr \end{matrix}\right]^{-1} 
\label{finalbis}
\end{eqnarray}
for configurations of the external states lying in
(\ref{condic})-(\ref{condic2}).

On the other hand, from (\ref{47}) we obtain
\begin{eqnarray}
 {\mathbb A}_4^{w=0}\left[
\begin{matrix}
j_1,j_2,j_3,j_4 \cr  j_1, m_2, -j_3,m_4 \cr \end{matrix}\right] 
&=& B(\tilde{j}_1) B(\tilde{j}_3) z^{m_2}
\overline{z}^{\overline{m}_2}\sum_{l=0}^s |z|^{2(\Delta_{\tilde{j}}-
\Delta_{\tilde{j}_1}-
\Delta_{j_2})} 
\nonumber \\
&& \times 
 A_3^{w=0}\left[
\begin{matrix}
\tilde{j}_1,j_2,\tilde{j} \cr  -\tilde{j}_1, m_2, -\tilde{m} \cr
\end{matrix}\right] 
A_3^{w=0}\left[
\begin{matrix}
\tilde{j}_3,j_4,\tilde{j} \cr  \tilde{j}_3,m_4,\tilde{m} \cr
\end{matrix}\right] 
A_2^{w=0}\left[
\begin{matrix}
\tilde{j},\tilde{j} \cr  -\tilde{m}, \tilde{m} \cr
\end{matrix}\right]^{-1},
\label{serserr2}
\end{eqnarray}
where $\tilde{j} = -1-\tilde{j}_1-j_2+l$.

Following the procedure leading to (\ref{47}) in the case of the
three-point functions, one can show from the Coulomb integral
expressions that
the factors $B(\tilde{j}_1)$ and $ B(\tilde{j}_3)$ can be reabsorbed 
as 
\begin{eqnarray}
&& B(\tilde{j}_1) A_3^{w=0}\left[
\begin{matrix}
\tilde{j}_1,j_2,\tilde{j} \cr  -\tilde{j}_1, m_2, -\tilde{m} \cr
\end{matrix}\right] 
= A_3^{w=1}\left[
\begin{matrix}
j_1,j_2,\tilde{j} \cr  j_1, m_2, -\tilde{m} \cr \end{matrix}\right], \nonumber
\end{eqnarray}
and a similar expression for the second three-point function in 
(\ref{serserr2}).
 Since the coordinate independent coefficient of the
two-point functions of states in different
spectral flow sectors does not change,
we finally get the
following expression:
\begin{eqnarray}
&&{\mathbb A}_4^{w=0}\left[
\begin{matrix}
j_1,j_2,j_3,j_4 \cr  j_1, m_2, -j_3,m_4 \cr \end{matrix}\right] 
\nonumber \\
&&~~~~
= \sum_{l=0}^s
|z|^{2(\hat{\Delta}_{\tilde{j},\tilde{m},{w}=-1}-
\Delta_{j_1}-
\Delta_{j_2})}A_3^{w=1}\left[
\begin{matrix}
j_1,j_2,\tilde{j} \cr  j_1, m_2, -\tilde{m} \cr \end{matrix}\right] A_3^{w=-1}\left[
\begin{matrix}
j_3,j_4,\tilde{j} \cr  -j_3, m_4, \tilde{m} \cr \end{matrix}\right] A_2^{w=0}\left[
\begin{matrix}
\tilde{j},\tilde{j} \cr  -\tilde{m}, \tilde{m} \cr \end{matrix}\right]^{-1},
\label{sees2}
\end{eqnarray}
the factor $z^{m_2}\,\overline{z}^{\overline{m}_2}$ in (\ref{serserr2})
being needed
in order to reproduce the correct conformal weight of the
intermediate states.

The equivalence between expansions of the same four-point function
in terms of  either $w$-conserving or
$w$-violating three-point functions
can be seen comparing Eqs. (\ref{serserr}) and (\ref{sees2}).

We mentioned above that the spectral flow makes the
validity of (\ref{final}) beyond the
semiclassical limit  
in the $AdS_3$
model 
more subtle 
than in the $H_3^+$
 model. But the possibility of encoding the
unflowed contributions in terms of 
spectral flowed intermediate states supports 
the conjecture that (\ref{final}) also holds for finite values of the affine
level \footnote{Additional indications that (\ref{final}) 
holds 
beyond the semiclassical limit
are given by the 
fact that the OPE leading to this expression in the bootstrap approach
to the $H_3^+$-WZNW model implemented in \cite{tesch1} reproduces the 
well-known fusion rules of admissible degenerate representations and
by the results on the structure of the factorization of string theory on 
$AdS_3$ in \cite{mo3}.}. 
If this is the case, starting from (\ref{finalbis}) 
instead of 
(\ref{serserr}) would lead us to the following analytic continuation of
(\ref{sees2}) \footnote{The analytic continuation leading to (\ref{finalbis})
cannot be directly 
implemented in the semiclassical limit for (\ref{sees2}).}
:
\begin{eqnarray}
&& {\mathbb A}_4^{w=0}\left[
\begin{matrix}
j_1,j_2,j_3,j_4 \cr  j_1, m_2, -j_3,m_4 \cr \end{matrix}\right] \nonumber \\
&&~~~~~~~~~ = \int_{\mathcal P}dj    ~
 |z|^{2(\hat{\Delta}_{\tilde{j},\tilde{m},\tilde{w}=-1}-
\Delta_{j_1}-
\Delta_{j_2})} \mathcal A_3^{w=1}\left[
\begin{matrix}
j_1,j_2,\tilde{j} \cr  j_1, m_2, -\tilde{m} \cr \end{matrix}\right] \mathcal A_3^{w=-1}\left[
\begin{matrix}
j_3,j_4,\tilde{j} \cr  -j_3, m_4, \tilde{m} \cr \end{matrix}\right] \mathcal A_2^{w=0}\left[
\begin{matrix}
\tilde{j},\tilde{j} \cr  -\tilde{m}, \tilde{m} \cr \end{matrix}\right]^{-1}. \nonumber
\label{finalconw}
\end{eqnarray}
Indeed, the equivalence of this last expression and 
(\ref{finalbis}) was obtained in
\cite{wc} and it was the starting point for an explicit verification that the
truncation imposed on the operator algebra by the spectral flow 
symmetry is realized at the level of physical amplitudes.

\section{Summary and discussion}

We have computed  $w-$conserving four-point correlation functions 
on the sphere in the 
$AdS_3$-WZNW model using the Coulomb gas approach. 
The requirement of integer numbers of screening operators,
a well known shortcoming of the formalism for models with continuous 
sets of fields, demands considering operators with
 quantized values of the
sum of their spins,
implying that only expectation values of
certain states in
 discrete representations can be evaluated without performing any
analytic continuation.
The result  in this case
was obtained as the monodromy invariant  sum of products of 
holomorphic and antiholomorphic  conformal blocks, namely, equation
(\ref{4pt}). 

The full integral expression for the conformal blocks presented in  (\ref{42})
 extends previous
results obtained
in \cite{dot2} where 
a simplified setting, sufficient to derive the operator algebra, 
was considered,
namely, two 
highest- and two lowest-weight operators with $j_1=j_4$ and $j_2=j_3$.
Indeed, we have computed
the
$\beta-\gamma$ ghost contributions 
for generic configurations of
fields, only restricted by the
 assumption of one highest-weight state and the
 requirement of arbitrary positive integer 
numbers
 of screenings. 
To this aim,
the  explicit computation of the ghost system involved in the
 three--point functions that was
presented in \cite{in}, although  not strictly necessary to obtain the
 Clebsch-Gordan coefficients, turned out to be a useful starting point 
to address
 the computation of
 these higher-point functions. Actually,
unlike the case of
the three-point functions, where
the full form of the kinematical
factor follows from the $SL(2)$ space-time symmetry of the model,
the ghost correlators give a non-trivial
dependence on the coordinates to the conformal blocks
of the four-point amplitudes. 

As discussed in section 2, 
relaxing the highest-weight condition 
assumed for one of the operators turns the elementary symmetric
 polynomials  in (\ref{integ}) into Schur polynomials and
 the coefficients given in (\ref{sc}) must be replaced by the
 most general expression  (\ref{pq}). Although the identities 
(\ref{223b}), (\ref{228b}) and (\ref{230b}) 
 that we have derived for the elementary symmetric polynomials
  cannot be easily extended to the
 general case, 
the $z, \overline z\rightarrow 0$ limit of the
integrals (\ref{212b}) has been computed in 
\cite{sergio} and then the
 conformal blocks for four generic states can be reconstructed 
along similar steps as those we have followed in subsection 2.3.
In any case, we have shown in section 3 that the leading terms
 in the factorization limit can be identified with products of
 three-point functions and then, from  (\ref{83b}) it is straightforward to
 see that the Selberg
 integrals in  (\ref{4pt}) must be replaced by combinations of
 Aomoto integrals in
the amplitude involving four  global descendants 
or their spectral flow images with spin values adding up
to an integer number.

Besides 
the ambiguities involved in the
 analytic continuation needed to obtain 
 amplitudes of arbitrary external states, 
the Coulomb gas method
suffers from the disadvantage of requiring quite a bit of  tedious algebra
 if compared to the bootstrap approach.
Nevertheless,
despite  these problems, we were able to present an alternative
derivation of the expression obtained for the four-point functions
 in \cite{tesch1, tesch3,
  tesch2}. Indeed, we have shown that 
for special configurations of fields,
 the semiclassical limit of the 
leading terms of the four-point function
may be rewritten as an integral over the spins of the intermediate states.
Actually, the  expression  (\ref{final}) obtained in section 3
 is valid for fields
 satisfying (\ref{condic}) and (\ref{condic2}) and
for other values of the kinematical parameters
it
must be defined by the
analytic continuation discussed in \cite{tesch3}, 
$i.e.$, the large-$k$ limit of the leading
terms of the amplitude are given by (\ref{final}) 
plus the contributions of all the poles that
cross the integration contour.  
This result reproduces in the $m-$basis the $x-$basis amplitude 
for  the $H_3^+$ model  obtained 
in \cite{tesch2} in the  mini-superspace 
approximation. 
The explicit Coulomb gas calculation that we have  presented here 
can thus be considered  as an independent check of  the
factorization ansatz based on the OPE of normalizable primary fields
 proposed in  \cite{tesch1, tesch3} for the $H_3^+$-WZNW model. 

The procedure followed in section 3 to
convert the discrete sums into the integral expression (\ref{final}) 
 provides a possible
resolution of the problem raised in \cite{dot2}
regarding the ambiguity involved in the analytic continuation of amplitudes 
containing
states with rational spin values in the $SU(2)$ CFT.
Moreover, it gives an
alternative route to the
use of the fractional calculus introduced in
\cite{rasmu}. Furthermore, since (\ref{final}) was  directly deduced
in the $m-$basis, it gives support to the process
implemented in \cite{wc, mn} to
transform  (\ref{4ptt}) from the $x-$basis  and helps clarify the related
questions raised in the introduction about exchanging the order of summation 
and
integration as well as convergence  of the
integral transform.

Following \cite{mo3, wc}, we have argued that the factorization into
 products of spectral flow preserving or violating 
three-point functions gives
 equivalent contributions to the $w$-conserving 
four-point functions and we proved
 this statement in a particular set of amplitudes using the
 Coulomb gas realization.
Indeed, we have shown  that the leading terms in the
factorization limit of the discrete sums
in  (\ref{4pt}) for certain amplitudes
can be rewritten alternatively
 as a sum of products  of two $w=0$ 
or of one $w=1$ and one $w=-1$ three-point functions.
 This result provides an 
independent confirmation of the
factorization $ansatz$ proposed in \cite{wc} and of the observation
 that $w$-conserving correlators in the
$AdS_3$-WZNW model are to be obtained from correlators in 
the $H_3^+$-WZNW model through
analytic continuation, although the spectral flow representations are not
 contained 
in the spectrum of the Euclidean  model.
Furthermore, it
 provides additional  support for the
 proposal
 that  (\ref{final})
also holds for finite $k$, beyond the semiclassical limit.

Not having achieved  a closed expression for the four-point functions
 cannot be attributed  to the  method.
Neither in the less complicated cases of Liouville theory or $H_3^+$-WZNW model
an explicit closed formula  is known. Although important progress has
recently been achieved in the former theory through the identification of the conformal
blocks  with Nekrasov's partition function of
certain ${\cal N}=2$ superconformal field theories \cite{alday,
  morozov}, 
the available amplitude is decomposed 
into structure
constants and $s-$channel conformal blocks that
have to
be numerically computed with
the techniques developed in
\cite{zamo}.
The existence of an interesting explicit formula for generic
four-point functions in the $AdS_3$-WZNW model also
seems unlikely.

It would be interesting 
 to extend the procedures developed in this paper to gain 
more insights into
 four-point functions and conformal blocks and to start understanding
 $w-$violating amplitudes in order to solve the $AdS_3$-WZNW model.
Indeed, there are several open problems yet.
The equivalence
between factorizations involving
$w$-conserving or $w$-violating three-point functions
  implies that the OPE (\ref{ope}) is actually equivalent to the OPE 
of normalizable states of the $H_3^+$-WZNW model proposed in 
\cite{tesch1, tesch3} when inserted into $w$-conserving amplitudes. 
But the latter OPE leads to
vanishing $w$-violating amplitudes, in contradiction with the 
spectral flow selection rules and the explicit computations performed in 
\cite{mo3, in}.
Moreover, the fusion rules
 obtained from the $H_3^+$-OPE by analytic continuation are not closed  in the
spectrum of the $AdS_3$-WZNW model and they are not compatible with the
identification
$\hat D_j^{\pm,w}=\hat D_{-k/2-j}^{\mp,w\pm 1}$ implied by the spectral
flow symmetry \cite{satoh, wc}. Furthermore, 
the OPE (\ref{ope}) has to be truncated
in order to avoid inconsistencies with the spectral flow symmetry and the
physical mechanism determining the decoupling is still not understood.
The computation of $w$-violating four-point functions might shed some light
on these issues. Although it requires the insertion of a spectral flow 
operator and consequently 
involves the evaluation of a five-point function, we hope to be able to 
address this problem in the near future
applying the techniques developed in this paper.

We expect that these techniques might also be useful to
 further deepen our knowledge on  properties of non-rational CFTs
and methods to deal with  them.

\bigskip

{\bf Acknowledgements:} We are grateful to  W. Baron, C. Cardona and P. Minces 
for reading the manuscript and many
stimulating discussions. It is our pleasure to thank the High Energy Group of the Abdus
Salam ICTP for the warm hospitality and stimulating atmosphere
during the completion of this
work. 
This work was partially supported by grants UBACyT X161
and PIP  CONICET 11220080100507.

\appendix

\section{Appendices}

\subsection{Evaluation of $m$-dependent coefficients: the general case}

For completeness, in this appendix we compute the coefficients $C_{nr}(z)$ in 
(\ref{cnp}) for
generic
 configurations of fields.

 Consider
the determinant of the matrix $(a_{ij})_{i,j=1}^{p}$ made up by
the first $p$ rows and $p$ columns of the matrix $(a_{ij}(z))_{i,j=1}^{s-n}$ 
with entries defined in
 (\ref{matrix}). Let us
denote it $d_p(z)$. As $d^0_p(z)$, 
this is a polynomial in $z$ of degree $p$ but satisfying
the following recurrence relation:
\begin{equation}
\label{recugrande}
d_p(z)=\ell_1^{s-n-r+p}(z) d_{p-1}(z) - \ell_2^{s-n-r+p-1}(z) 
\ell_0^{s-n-r+p}(z)
d_{p-2}(z),
\end{equation}
with boundary conditions
$d_1(z)=\ell_1^{s-n-r+1}(z)$ and
$d_2(z) = \ell_1^{s-n-r+1}(z) \ell_1^{s-n-r+2}(z) - \ell_2^{s-n-r+1}(z) \ell_0^{s-n-r+2}(z) $. To deduce (\ref{recugrande}) we have used that $(a_{ij}(z))_{i,j=1}^{p}$ is a tridiagonal
matrix.

It is convenient to introduce the following ``shifted'' parameters:
\begin{equation}
\left\{
\begin{array}{l}
	\alpha'= -s+n+r+\alpha \\
	\beta'= -s+n+r+\alpha+\beta \\
	\gamma'= -s+n+r+\alpha+\gamma,
\end{array}
\right. \nonumber
\end{equation}
and rewrite (\ref{recugrande}) more explicitly as
\begin{eqnarray}
\label{recunuec}
&& d_p(z) = \left [(1-p+\beta') + (1-p+\gamma')z \right ] d_{p-1}(z) -
(p-2-\beta'-\gamma')
(p-1) z d_{p-2}(z)\nonumber\\
&&~~~~~~~~~~~~~~~~~~~~~~~~~+
 \alpha' (\alpha'-\beta'-\gamma'-1) z d_{p-2}(z),
\end{eqnarray}
with $d_1(z)=\beta' + \gamma'z$ and $
d_2(z) = \beta'(\beta'-1) + 2 \beta'\gamma' z + \gamma'(\gamma'-1)z^2 +
\alpha'(\alpha'-\beta'-\gamma'-1)z$.

For the case $\alpha '=0$ we have found the solution of this
recurrence in (\ref{sum}), namely,
\begin{equation}
d^0_{p}(z)= \frac{\Gamma(\beta'+1)}{\Gamma(\beta' - p + 1 )} \,
{}_2F_1\left [ \left .\begin{array}{c}
-p,-\gamma' \\ \beta'-p+1 \end{array} \right | z \right ]. \nonumber
\end{equation}

Defining $d_p(z)=d^0_p(z)+\epsilon_p(z)$ and replacing this into 
(\ref{recunuec})
we obtain the following recurrence for $\epsilon_p(z)$:
\begin{eqnarray}
\epsilon_p(z) &=& \left [(1-p+\beta') + (1-p+\gamma')z \right ] 
\epsilon_{p-1}(z)
 \nonumber \\
&-&
(p-2+\alpha'-\beta'-\gamma') (p-1-\alpha') z \epsilon_{p-2}(z) + 
\alpha' (\alpha'-\beta'-\gamma'-1) z d^0_{p-2}, \nonumber
\end{eqnarray}
with
$\epsilon_1(z)=0$ and
$\epsilon_2(z) =\alpha'(\alpha'-\beta'-\gamma'-1)z$.

Inductively solving this much simpler recurrence it is possible to show that
\begin{equation}
\label{37}
d_{p}(z) = \sum_{t=0}^{[p/2]} \left(
\begin{array}{c}
	p-t \\ t
\end{array}
\right) \frac{\Gamma(\alpha'+1) \Gamma(\alpha'-\beta'-\gamma'+t-1)}{\Gamma(\alpha'-t+1)
\Gamma(\alpha'-\beta'-\gamma'-1)} d^{[t]}_{p-2t}(z),
\end{equation}
where we have defined
\begin{equation}
d_p^{[t]}(z)=\frac{\Gamma(\beta'-t+1)}{\Gamma(\beta'-t - p + 1 )} \,
{}_2F_1\left [ \left .\begin{array}{c}
-p,-\gamma'+t \\ \beta'-t-p+1 \end{array} \right | z \right ].\nonumber
\end{equation}

Therefore, the correlation function involving four generic states is given by
\begin{eqnarray}
\label{1}
A_4^{w=0}\left[
\begin{matrix}
j_1,j_2,j_3,j_4 \cr  m_1, m_2, m_3,m_4 \cr \end{matrix}\right]
 =  \Gamma(-s) |z|^{4 j_1 j_2 \rho} |1-z|^{4 j_2 j_3 \rho}
\sum_{n,\overline{n}=0}^{s} \sum_{r,\overline{r}=0}^{s-n} \left[C_{nr}(z)
\times \, c.c.\right] {\cal J}_{nr,\overline{n}
\overline{r}}(z,\overline{z}),
\end{eqnarray}
with
\begin{eqnarray}
C_{nr}  &=&   (-1)^{s-n-r} \frac{\Gamma(\alpha+1)}{\Gamma(\alpha
  -s+n+r+1)}
\frac{\Gamma(s-\alpha-\beta-\gamma)}{\Gamma(s-n-\alpha-\beta-\gamma)}\,
\nonumber\\
&\times &\sum_{t=0}^{\infty} \left(
\begin{array}{c}
	s-n-t \\ t
\end{array}
\right) \frac{\Gamma(\alpha'+1) \Gamma(\alpha'-\beta'-\gamma'+t-1)}{\Gamma(\alpha'-t+1)
\Gamma(\alpha'-\beta'-\gamma'-1)} d^{[t]}_{s-n-2t}(z)\,  z^{s-n-r}
\label{pq}
\end{eqnarray}
where we have used, again, that the sum in (\ref{37}) can be set to $\infty$.

Recall that this expression strictly corresponds to a correlator with integer
number of
screening operators, {\em i.e.},
$s=j_1+j_2+j_3+j_4+1 \in \mathbb{N}_0$.
The four-point function for general kinematic configurations
 is assumed to be given by an analytic continuation of (\ref{1})
for non-integer values of $s$, as discussed in section 3.

\subsection{Poles of the three-point function and the reflection symmetry}

Let us assume that the three-point function ${\mathcal A}_3^{w=0}\left[
\begin{matrix}
j_1,j_2,j_3 \cr  m_1, m_2, m_3 \cr \end{matrix}\right]$ has a pole located at $j_3 = f(j_1,j_2)$. It is clear that it could also have a pole at $j_3 = f(j_1,-1-j_2)$ and that
\begin{eqnarray}
\mbox{Res}_{j_3=f(j_1,-1-j_2)} {\mathcal A}_3^{w=0}\left[
\begin{matrix}
j_1,j_2,j_3 \cr  m_1, m_2, m_3 \cr \end{matrix}\right] = \left\{ \mbox{Res}_{j_3=f(j_1,j_2)}{\mathcal A}_3^{w=0}\left[
\begin{matrix}
j_1,-1-j_2,j_3 \cr  m_1, m_2, m_3 \cr \end{matrix}\right]\right\}_{j_2 \rightarrow -1-j_2}.
\label{prim}
\end{eqnarray}

From \cite{satoh} we know that
\begin{eqnarray}
 {\mathcal A}_3^{w=0}\left[
\begin{matrix}
j_1,-1-j_2,j_3 \cr  m_1, m_2, m_3 \cr \end{matrix}\right]  = 
B(j_2)\,c^{j_2}_{m_2,\overline{m}_2} {\mathcal A}_3^{w=0}\left[
\begin{matrix}
j_1,j_2,j_3 \cr  m_1, m_2, m_3 \cr \end{matrix}\right]. \nonumber
\end{eqnarray}
Inserting this expression into (\ref{prim}) we obtain
\begin{eqnarray}
\mbox{Res}_{j_3=f(j_1,-1-j_2)} {\mathcal A}_3^{w=0}\left[
\begin{matrix}
j_1,j_2,j_3 \cr  m_1, m_2, m_3 \cr \end{matrix}\right] 
 = B(-1-j_2)\,c^{-1-j_2}_{m_2,\overline{m}_2}\left\{
 \mbox{Res}_{j_3=f(j_1,j_2)} 
{\mathcal A}_3^{w=0}\left[
\begin{matrix}
j_1,j_2,j_3 \cr  m_1, m_2, m_3 \cr \end{matrix}\right]\right\}_{j_2 \rightarrow -1-j_2}.
\nonumber
\end{eqnarray}
This equation displays the relation between the residues of the
three-point function 
associated to poles related by reflection.

If the state inserted at $z_2,\overline{z}_2=1$ lies in a discrete
series, it 
follows that ${\mathcal A}_3^{w=0}\left[
\begin{matrix}
j_1,j_2,j_3 \cr  m_1, m_2, m_3 \cr \end{matrix}\right]$ is actually
regular at 
$j_3=f(j_1,-1-j_2)$ since in this case $c^{-1-j_2}_{m_2,\overline{m}_2}$ vanishes.

\end{document}